\documentclass[showpacs,showkeys,preprint,amsmath,amssymb,nofootinbib, prd, aps, 11pt]{revtex4-1}
\usepackage[T1]{fontenc}
\usepackage[latin1]{inputenc}

\usepackage[sc]{mathpazo}
\usepackage{mathrsfs}
\usepackage{framed, color}
\definecolor{shadecolor}{rgb}{0.84,0.84,0.84}
\usepackage{graphicx}
\newcommand{\tr}{\mbox{tr}}

\newcommand{\Vek}[1]{\mbox{\boldmath$#1$\unboldmath}}
\newcommand{\vek}[1]{\mathbf{#1}}

\newcommand{\QQ}{Q}
\newcommand{\ab}{\mathsf{a}}
\newcommand{\db}{\mathsf{d}}

\newcommand{\ww}{\bar{\omega}}

\newcommand{\dd}{\text{\dj}}

\newcommand{\xx}{\bar{\chi}}

\begin{document}


\title{Covariant variational approach to Yang-Mills Theory: effective potential of the Polyakov loop}

\author{M.~Quandt}\email{markus.quandt@uni-tuebingen.de}
\author{H.~Reinhardt}\email{hugo.reinhardt@uni-tuebingen.de}
\affiliation{%
Universit\"at T\"ubingen\\
Institut f\"ur Theoretische Physik\\
Auf der Morgenstelle 14\\
D-72076 T\"ubingen, Germany
}%

\date{\today}


\begin{abstract}
We compute the effective action of the Polyakov loop in $SU(2)$ and $SU(3)$
Yang-Mills theory using a previously developed covariant variational approach. 
The formalism is extended to background gauge and it is shown how to relate the 
low order Green's functions to the ones in Landau gauge studied earlier. The 
renormalization procedure is discussed. The self-consistent 
effective action is derived and evaluated using the numerical solution of the 
gap equation. We find a clear signal for a deconfinement phase transition at 
finite temperatures, which is second order for $SU(2)$ and first order for 
$SU(3)$. The critical temperatures obtained are in reasonable agreement 
with high precision lattice data.
\end{abstract}


\maketitle


\section{Introduction}
\label{intro}
The low energy sector of quantum chromodynamics (QCD) and, in particular, 
its phase diagram continue to be one of the most actively researched topics 
in elementary particle physics. Recently, large experimental facilities such
as the large hadron collider (LHC) offer the possibility to study strongly 
interacting matter under extreme conditions, such as large temperatures and 
baryon densities. On the theoretical side, lattice simulations can be used
to obtain numerical 
\emph{ab-initio} solutions of QCD in a variety of settings, but restricted to 
zero or small chemical potential due to the \emph{sign problem}. 
Alternative continuum or functional methods, which are not plagued by this issue, 
are thus of particular intereset. These methods comprise the functional 
renormalization group (FRG) flow equations \cite{Pawlowski:2005xe,Gies:2006wv}
Dyson-Schwinger equations (DSE) \cite{Fischer:2006ub,Alkofer:2000wg,Binosi:2009qm} 
and variational methods \cite{Quandt:2013wna,Feuchter:2004mk,Feuchter:2004gb}. 
In all of these approaches, 
the simplest objects to be studied are the low-order Green's functions of
gluons, ghosts and quarks. Since these quantities are, however,  gauge-dependent, 
one has to fix a gauge in order to get meaningful results. The DSE and FRG 
approach have a covariant setup and are thus predominantly studied in Landau gauge, 
while the variational approach has been formulated and worked out in the 
Hamiltonian approach using Coulomb gauge. 

In the covariant case, the Becchi-Rouet-Stora-Tyutin (BRST) symmetry can be 
used to guide the analsysis, and the Kugo-Ojima criterion \cite{Ojima:1978hy,Kugo:1979gm} 
claims a  direction connection (based on an unbroken BRST symmetry) between the 
propagators in Landau gauge and color confinement. In reality, however, 
the functional methods in Landau  gauge allow for a whole family of solutions 
for the propagator which are parametrized by different infrared 
boundary conditions \cite{Fischer:2008uz}. In particular, the so-called \emph{decoupling solution} 
shows excellent agreement with the lattice data, even though it (softly) breaks BRST 
symmetry. The Kugo-Ojima criterion is no longer applicable in this case, and 
the question of color confinement must be resolved in other ways. One particularly 
simple approach is to study the effective action for the \emph{Polyakov loop}, a 
maximally extended temporal Wilson line that closes via the periodic boundary conditions 
in the euclidean time direction. The Polyakov loop is a direct order parameter for 
confinement \cite{McLerran:1981pb, Svetitsky:1985ye} and can therefore be used to study the 
\emph{deconfinement phase transition} 
that occurs in the QCD phase diagram at increasing temperatures. 
Naturally, the FRG \cite{Braun:2007bx, Marhauser:2008fz, Braun:2012zq}, 
the DSE \cite*{[{see e.g.~}][{ and references therein}]Luecker:2013oda}
and the Hamiltonian approach in 
Coulomb gauge \cite{Reinhardt:2012qe, Heffner:2012sx, Reinhardt:2013iia} have been used to 
study the properties of this phase transition 
and compared it with findings from high precision lattice studies. In the pure Yang-Mills 
case, both approaches predict a transition which is second order for the color group $G=SU(2)$ 
and first order for $G=SU(3)$. Moreover, the phase transition temperatures are in 
reasonable quantitative agreement with the lattice. 

In Refs.~\cite{Quandt:2013wna} we have proposed an alternative covariant variational approach which 
combines the theoretical simplicity of the covariant setup with the efficiency of 
a variational method. It yields a closed set of integral equations that can be 
renormalized by conventional counter-term techniques. In addition, the variational 
character automatically optimizes any ansatz for the low-order Green's functions, which 
therefore are predicted quite accurately even when based on simple Gaussian ans\"atze for the 
trial measure. In particular, the quantitative agreement with the lattice propagators 
at $T=0$ is excellent \cite{Quandt:2013wna}, and also the corrections due to  finite temperatures are
reproduced in accord with the lattice data \cite{Quandt:2015aaa}. 
Due  to the inherent breaking of BRST symmetry, the Kugo-Ojima ciriterion does not 
apply and the question of color confinement is inconclusive. As mentioned above, 
it is therefore important to study the effective potential of the Polyakov loop to see 
if our method, combined with the simple Gaussian ansatz, is strong enough to predict 
the properties of the QCD phase diagram in this region.  In the present paper, we study 
this question at vanishing chemical potential and for pure Yang-Mills theory, i.e.~in the 
absence of dynamical quarks. Both these restrictions are intended to be lifted in 
future studies. 

Our paper is organized as follows:  In section \ref{center}, we recall the physical 
interpretation of the Polyakov loop and its relation to center symmetry and to color 
confinement. Section \ref{back} extends our variational approach to  the background 
gauge which is most convenient for the study of the Polyakov loop in our setup.  
In particular, the relevant variational kernels need not be re-computed, but instead
are inherited from the known kernels in Landau gauge, as is demonstrated in section 
\ref{relating}. The following section \ref{EA} computes the self-consistent effective 
action for the Polyakov loop in our approach and shows how to turn these results in 
an efficient numerical calculation. Section \ref{numerics} describes our numerical 
setup and presents our results, including a comparision with high-precision lattice
simulations. Finally, we conclude in section \ref{conclusions} with a short 
summary and an  outlook on the future development of our approach. 
 
\section{Center symmetry and the Polyakov loop}
\label{center}
Let us recall the role of the center symmetry and the Polyakov loop
as an order parameter for the deconfinement phase transition in pure Yang-Mills
theory. At finite temperatures, the coupling to the heat bath excites higher
energy states and the static quark potential $V_\beta(r)$ at inverse temperatuer $\beta$ 
must be defined as the change in \emph{free energy} due to a heavy $q\bar{q}$-pair 
immersed in the thermal background \cite{Svetitsky:1985ye}. 
Within the imaginary time formalism, $V_\beta(r)$ can extracted from non-contractible 
\emph{Polyakov loops} winding around the compactified Euclidean time direction:
\begin{align}
\mathsf{L}(\mathbf{x}) = \mathrm{tr}\,\mathsf{P}\exp\left[ - \int_0^\beta A_0(\tau,\vek{x}) d\tau
\right]\,,
\end{align}
where $\mathsf{P}$ is path ordering and the trace in the respective quark representation is 
normalised such that $\mathrm{tr} \mathbb{1} = 1$. In a pure Yang-Mills theory, the insertion of 
a static heavy quark at position $\vek{x}$ in the thermal background state is, in fact, equivalent 
to the insertion of a Polyakov loop in the finite-temperature partition function \cite{McLerran:1981pb}. 
More precisely, the free energy of a static $q\bar{q}$-pair located at positions $\vek{x}$ and 
$\vek{y}$ is obtained from the correlator of two Polyakov lines, 
\begin{align}
\exp\big[- \beta \,V_\beta(r)\big] = Z^{-1}(\beta)\,\mathrm{Tr}_{\rm phys}\Big[
\mathsf{L}(\vek{x})\cdot L^\ast(\vek{y})\cdot e^{-\beta H_{\rm phys}} \Big] 
\equiv \big\langle\,\mathsf{L}(\vek{x})\cdot \mathsf{L}^\ast(\vek{y})\,\big\rangle\,,
\label{x2}
\end{align}
where $r = | \vek{x}-\vek{y} |$ is the distance of the quarks and the star on $\mathsf{L}$ 
denotes complex conjugation. Furthermore, $H_{\rm phys}$ is the gauge invariant 
Yang-Mills Hamiltonian and the partition function $Z(\beta)$ serves to normalize the 
expectation value. Since the pure phases on both sides of the deconfinement 
transition are expected to show asymptotic clustering, we find
\begin{align}
\big\langle\, \mathsf{L}(\vek{x})\cdot \mathsf{L}^\ast(\vek{y})\,\big\rangle 
\to \big\langle\,\mathsf{L}(\vek{x})\,\big\rangle\cdot \big\langle\,\mathsf{L}^\ast(\vek{y})\,\big\rangle
= \big\vert\big\langle\,\mathsf{L}\,\big\rangle\big\vert^2\qquad\quad\text{as}\quad 
|\vek{x}-\vek{y}| \to \infty\,.
\label{x3}
\end{align}
From eqs.~(\ref{x2}) and (\ref{x3}), we conclude that a vanishing average for the Polyakov line 
immediately leads to $\big\langle\, \mathsf{L}(\vek{x})\cdot \mathsf{L}^\ast(\vek{y})\,\big\rangle \to 0$
for large $q\bar{q}$ separations, i.e.~a static potential increasing with the distance between the 
charges.
Thus, $\big\langle L \big\rangle = 0$ implies color confinement, while $\big\langle L \big\rangle\neq 0$ 
implies that the static $q\bar{q}$-potential remains finite, and a finite energy is sufficient 
to separate the $q\bar{q}$-pair (\emph{deconfinement}).\footnote{The formal association of a 
free energy of a fictious single quark, $\exp(- \beta F_q) = \big\langle \mathsf{L} \big\rangle$, 
is, however, doubtful: since a single color charge in the fundamental representation cannot be 
screened by gluons, Gauss' law cannot be satisfied within the vacuum sector of gauge-invariant states,
and the physical traces are void.} 
The confined phase is characterized as an unordered phase with a higher degree of 
symmetry. In fact, the vanishing of the Polyakov loop can be seen as a consequence of 
\emph{center symmetry} which maps each Polyakov loop $\mathsf{L} \to z\,\mathsf{L}$ by a 
center element $z$ of the color group, but leaves the action invariant. 
If unbroken, $\langle \mathsf{L} \rangle = z_k \,\langle \mathsf{L}\rangle$ for each 
center element $z_k$, and thus $\langle \mathsf{L}\rangle = 0$ because $\sum_k z_k = 0$.
An unbroken center symmetry hence implies confinement. Conversely, a broken center symmetry 
implies $\langle \mathsf{L} \rangle \neq z_k \,\langle \mathsf{L} \rangle$ for each center 
element, and thus $\langle \mathsf{L} \rangle \neq 0$, which entails deconfinement. 

In a lattice discretisation, the center symmetry can be realized by multiplying all 
temporal links emenating from a fixed time slice by the same center element $z$. 
A continuum version of this construction employs the so-called \emph{Polyakov gauge},
\begin{align}
\partial_0 A_0^{a'} = 0\,,\qquad\qquad A_0^{\bar{a}} = 0\,,
\end{align}
where $\{T^{a'}\}$ are a maximal set of commuting generators of the Lie algebra of the 
color group $G$, which span the Cartan subgroup (maximal torus) $H \subset G$, while 
the remaining generators span the coset $G/H$. The Polyakov loop becomes
\begin{align}
\mathsf{L}(\mathbf{x}) = \mathrm{tr}\,\exp\left[ - \beta A_0^{a'}(\vek{x}) T^{a'}\right]
\end{align}
and requires no path ordering. For the color group $G=SU(2)$, the maximal torus is spanned 
by $T^3 = \sigma^3 / (2i)$ (where $\sigma^3$ is the diagonal Pauli matrix) and the Polyakov 
loop can be expressed as 
\begin{align}
\mathsf{L}(\vek{x}) = \cos \big(\pi \phi(\mathbf{x})\big)\,,\qquad\qquad 
\phi(\mathbf{x}) =   \frac{\beta}{2\pi}\,A_0^3(\vek{x})\,.
\end{align}
The fundamental modular region is $\phi(\mathbf{x}) \in [0, 1]$ and the effective 
potential for the observable $\phi$ is even in $\phi$ (due to Weyl reflection symmetry)
and periodic outside the range $[-1,1]$.  The \emph{center symmetry} changes 
$\mathsf{L}(\vek{x}) \to  -\mathsf{L}(\vek{x})$ or $\phi(\vek{x}) \to 1 - \phi(\vek{x})$ at 
\emph{all} space positions simultaneously. It is a global symmetry which may be 
broken or preserved dynamically. In any event, 
the effective potential $\Gamma[c]$ for the Polyakov loop $c = \langle \cos(\pi\phi)\rangle$
will be symmetric under center transformation, $\Gamma[-c] = \Gamma[c]$ and the symmetry 
breaking only dictates whether the minimum is at $c=0$ or at the two pure phases 
$c = \pm 1$ (or in between in a mixed phase). It is therefore sufficient to study $\Gamma[c]$
in the range $c = \langle \cos(\pi\phi)\rangle \in [0,1]$
and recover the other half of the effective potential by center symmetry.

In Refs.~\cite{Marhauser:2008fz} and \cite{Braun:2007bx}, it was argued that not only the 
Polyakov loop, but also the expectation value of its logarithm 
(i.e. $\langle \phi \rangle \sim \langle A_0^3\rangle$)
can be an order parameter for confinement. Let us briefly repeat 
this reasoning: In the \emph{confined phase}, the traced Polyakov loop vanishes and 
center symmetry is unbroken. This implies that configurations $\phi(\vek{x})$ and 
$1 - \phi(\vek{x})$ are equally likely, 
i.e.~$\langle \phi(\vek{x})\rangle = 1 - \langle \phi(\vek{x})\rangle$ 
or $\langle \phi(\vek{x}) \rangle = 1/2$ at any space position $\vek{x}$. 
As a consequence
\begin{align}
\text{confined (center symmetric):} \qquad\langle \mathsf{L} \rangle 
= \langle \cos(\pi \phi) \rangle = 0 = \cos\big(\pi\langle\phi\rangle\big)\qquad
\text{i.e.}\quad \langle \phi \rangle = \frac{1}{2}\,.
\end{align}
In the \emph{deconfined phase}, $\langle \mathsf{L} \rangle = \langle \cos (\pi \phi) \rangle > 0$
within the fundamental domain, and hence $\langle | \cos (\pi\phi) | \rangle > 0$. (The latter 
conclusion follows because 
$\langle | \cos (\pi\phi) | \rangle = 0$ would imply $| \cos (\pi\phi) | = 0$ almost everywhere, hence 
$\cos (\pi\phi) = 0$ almost everywhere and thus $\langle \cos (\pi\phi) \rangle = 0$ in contradiction 
to the assumption.) Since the function $\phi \mapsto | \cos(\pi \phi) |$ is concave 
on the fundamental domain, Jensen's inequality can be applied 
\cite{Marhauser:2008fz, Braun:2007bx} and we have 
\begin{align}
\text{deconfined (center broken):} \qquad 0 < \langle | \cos(\pi\phi) | \rangle \le | 
\cos(\pi\langle \phi \rangle)|
\qquad\text{i.e.}\quad \langle \phi \rangle \neq \frac{1}{2}\,.
\end{align}
Thus, for the gauge group $G=SU(2)$ at least, we can also use
$\langle \phi \rangle \sim \langle A_0 \rangle$ itself as an order parameter for 
confinment in Polyakov gauge.\footnote{For arbitrary color groups, the corresponding inequality 
$0 < \langle \mathsf{L}[A_0]\rangle \le \mathsf{L}[\langle A_0 \rangle]$ is invalid. 
In Ref.~\cite{Braun:2012zq}, it was argued that it holds at least in the special 
case $G=SU(3)$ at any temperature, and for arbitrary color groups at sufficiently high 
temperatures.} 

\medskip
The considerations above were valid in the (non-covariant) Polyakov gauge. Since the 
effective action for a gauge-variant operator such as $A_0$ is also gauge-variant,
$\Gamma[\langle \phi \rangle]$ will differ when changing gauges. For instance, 
$\langle \phi \rangle \sim \langle A_0 \rangle = 0$ in Landau gauge due to Lorentz and 
global color invariance, and nothing can be learned from $\langle A_0 \rangle$. In order 
to have a covariant theory with $\langle A_0 \rangle \neq 0$, we have to adopt the 
\emph{background gauge formalism} described below. Even then, the effective action for 
the background field $\mathsf{a}_0 = \langle A_0 \rangle$ does not necessarily agree 
with the one in Polyakov gauge discussed above. However, it has been argued in 
Refs.~\cite{Marhauser:2008fz, Braun:2007bx} that the background gauge formalism
\emph{can} be used to detect gauge invariant features such as the location and the order 
of the phase transition, if the background field $\mathsf{a}_\mu$ itself is taken to 
have only an $\mathsf{a}_0$ component which obeys the Polyakov gauge condition.
Moreover, the corresponding potentials in background and Polyakov 
gauge are very similar numerically, both on the lattice and in the functional 
renormalization group approach \cite{Marhauser:2008fz}. In the latter case, it was argued that
this coincidence might be expected, as the relevant quantum fluctuations in background gauge 
include the ones in Polyakov gauge, \emph{if} the backgrond field $\mathsf{a}_0$ itself is 
taken in Polyakov gauge.\footnote{Recent calculations for the gauge-invariant 
Polyakov loop directly seem to indicate that the phase transition temperature might be 
somewhat higher as compared to the findings for $\langle A_0 \rangle$ in background gauge 
\cite{Herbst:2015ona}.}
We will therefore adopt this formulation and employ the 
covariant variational approach \cite{Quandt:2013wna,Quandt:2015aaa} to compute 
the effective action for a time-independent, Abelian background field $\mathsf{a}_0(\mathbf{x})$, and check if the 
formalism is strong enough to predict the known properties of the deconfinement phase 
transition for $G=SU(2)$ and $G=SU(3)$.

\section{The covariant variational approach in background gauge}
\label{back}
In the background field formalism, we choose an arbitrary background field $\ab_\mu$ and 
decompose the full gauge connection $A_\mu = \ab_\mu + \QQ_\mu$, where $\QQ_\mu$ is the 
fluctuating field. We fix the quantum gauge symmetry ($\theta = $ infinitesimal gauge angle)
\begin{align}
 \delta \ab_\mu = 0\,,\qquad\quad \delta \QQ_\mu = [ D_\mu(\QQ)\,,\,\theta] + [\ab_\mu\,,\, \theta]
\end{align}
by imposing the background g.f.~condition
\begin{align}
[\db_\mu\,,\,Q_\mu] = 0\,,
\label{bgg}
\end{align}
where $\db_\mu \equiv D_\mu(\ab)$ is the covariant background deriative.
It is then easy to see that the gauge-fixed action, the corresponding path integral and 
hence the quantum effective action $\Gamma_\ab[\mathcal{Q}]$ for the 
classical fluctuation field $\mathcal{Q}_\mu = \langle Q_\mu \rangle_{\mathsf{a}} $ 
(with the fixed background field $\ab_\mu$ as a parameter) are all invariant under the 
background transformation
\begin{align}
 \delta \ab_\mu =  [\db_\mu\,,\,\theta]\,,\qquad\quad \delta \QQ_\mu = [\QQ_\mu\,,\, \theta]\,.
\end{align}
Also $\Gamma_\ab[\mathcal{Q}] = \Gamma[\ab + \mathcal{Q}]$, where 
$\Gamma$ is now the usual effective action, i.e.~the generating functional of proper functions
for the \emph{original} gauge field $A_\mu$, in the (slightly unusual) gauge
\begin{align}
[ \db_\mu\,,\,A_\mu - \ab_\mu] = 0\,. 
\label{komisch_gauge}
\end{align}
If we set the classical field $\mathcal{Q}_\mu = 0$, we find that 
$\Gamma[\ab] = \Gamma_\ab[0]$ is a gauge-invariant functional of the background field $\ab_\mu$.

\medskip
Next, we have to adapt this formalism to the variational approach of 
Refs.\cite{Quandt:2013wna,Quandt:2015aaa}. We keep $\ab_\mu$ as a parameter and define the 
variational approach to the path integral for the fluctuation field, 
\begin{align}
\Gamma_\ab[\mathcal{Q}] = \min_\mu \Big\{ 
F_\ab(\mu) \equiv \langle S_{\rm fix} \rangle_\mu - \hbar \overline{W}(\mu)\,\vert\,
\langle Q \rangle_\mu = \mathcal{Q}\Big\}\,.
\label{var}
\end{align}
Here, the variation is over all normalised path integral measures $\mu$ for the 
fluctuation field, $S_{\rm fix}$ is the Yang-Mills action with the gauge fixing term 
corresponding to the background gauge condition (\ref{bgg}), and 
\begin{align}
\overline{W} \equiv  - \big \langle \,\ln(\rho/\mathcal{J})\,\big\rangle_\mu
\end{align}
is the entropy of the measure $d\mu \equiv dQ\,\rho(Q)$ relative to the natural measure on 
the gauge orbit \cite{Quandt:2013wna}, i.e.~the Faddeev-Popov determinant $\mathcal{J}$.
Before investigating this further, it should be stressed once again that the effective action 
eq.~(\ref{var}) is \emph{not} the same as the effective action $\Gamma[A]$ for the original 
gauge field $A_\mu$ (in the unusual gauge eq.~(\ref{komisch_gauge})), because the \emph{vev} 
of the fluctuation $Q_\mu$, rather than the full $A_\mu$, is held fixed. As mentioned earlier, 
$\Gamma_\ab[\mathcal{Q}] = \Gamma[\ab + \mathcal{Q}]$ 
and we set the \emph{classical} fluctuation field $\mathcal{Q}_\mu = 0$ from this point on. 
In particular, this means that all trial measures for the variation principle (\ref{var})  
must be centered in the origin, $\langle Q_\mu \rangle = 0$.

\bigskip
To further work out the variational approach, we write the measure $\mu$ as a modification of the 
flat measure, $d\mu(Q) = dQ\,\rho(Q)$ and denote the Faddeev-Popov determinant in background gauge by
\begin{align}
\mathcal{J}(A) = \mathcal{J}(\ab + Q)\equiv \det\Big[- \hat{\db}_\mu \hat{D}_\mu(\ab + Q)\Big]\,, 
\label{FP}
\end{align}
where the hat denotes the adjoint representation of the color group, 
i.e.~$\hat{D}_\mu^{ab} = \partial_\mu\,\delta^{ab} + \hat{A}_\mu^{ab} = 
\partial_\mu\,\delta^{ab} - f^{abc} A_\mu^c$ with the antisymmetric structure 
coefficients $f^{abc}$, and likewise for the covariant background derivative 
$\hat{\db}_\mu = \hat{D}_\mu(\ab)$. Then the free action $F(\mu)$ for a measure $\mu$ is 
given explicitly by
\begin{align}
F_\ab(\mu) = \langle S_{\rm YM}(\ab + Q)\rangle_\mu + \frac{\xi}{2}
\left\langle \int \big[\db_\mu, Q_\mu]^2 \right\rangle_\mu - \langle \ln \,\mathcal{J}\rangle_\mu 
+ \langle \ln \rho \rangle_\mu\,,
\label{free}
\end{align}
where $\xi$ is a gauge-fixing parameter. We cannot do the variation over the full set of 
measures $d\mu(Q)$. Instead, we will restrict our investigations to \emph{Gaussian} trial measures, 
since these have already proven to be successful in the description of the low-order  
Green's functions, both at zero \cite{Quandt:2013wna} and finite temperature \cite{Quandt:2015aaa}. 
More specifically, our ansatz is
\begin{align}
d\mu &=  \mathcal{N}\cdot \mathscr{D}A\,\mathcal{J}(A)^{1-2\alpha}\,\exp\left[ - \frac{1}{2}\,
\int d(x,y)\,(A - \ab)^a_\mu\,\omega_{\mu\nu}^{ab}(x,y)\,(A - \ab)^b_\nu(y)\right]
\nonumber \\[2mm]
&= \mathcal{N}\cdot \mathscr{D}Q\,\mathcal{J}(\ab + Q)^{1-2\alpha}\,\exp\left[ - \frac{1}{2}\,
\int d(x,y)\,Q^a_\mu(x)\,\omega_{\mu\nu}^{ab}(x,y)\,Q^b_\nu(y)\right]\,,
\label{ansatz}
\end{align}
where $\omega$ is a variational kernel to be determined by minimizing the free action 
eq.~(\ref{free}). Furthermore, the  Gaussian is centered at $\langle A_\mu \rangle = \ab_\mu$ 
(or $\langle Q_\mu \rangle = 0$) as discussed above. Note that the FP determinant in the 
first line is evaluated at the original field $A$ as in eq.~(\ref{FP}), since this is the natural measure 
on the original gauge orbit. After the shift to the fluctuation field in the second step,  the measure 
remains centered, $\langle Q_\mu \rangle = 0$, even when the FP prefactor is taken into account. This 
is clear because fluctuations about the background field with opposite signs should be equally likely 
on the gauge orbit, and it will be verified explicitly when $\mathcal{J}$ is treated in curvature 
approximation \cite{Reinhardt:2004mm}, cf.~eq.~(\ref{curvxx}) below. A side effect of the curvature 
approximation is that the variational parameter $\alpha$, which controls the localization of the measure near 
the Gribov horizon, becomes immaterial and the trial measure is exactly Gaussian for all $\alpha$. 
This is benificial, because it allows to use Wick's theorem to evaluate expectation values. Since the 
free action (\ref{free}) depends implicitly on the background field $\ab$, we must allow for the 
variational kernel $\omega$ in the ansatz (\ref{ansatz}) to be non-transversal and 
non-diagonal in color. 

Next we insert the ansatz (\ref{ansatz}) in the free action eq.~(\ref{free}). After a short calculation, 
we find
\begin{align}
F[\ab] = F_\ab[\mathcal{Q}=0] = \langle S_{\rm YM}[\ab + Q]\rangle_\omega + \frac{\xi}{2}\,
\left\langle \int \big(\hat{\db}_\mu\,Q_\mu\big)^2\right\rangle_\omega - 2 \alpha \,\langle \ln \mathcal{J}\rangle_\omega 
+ \ln \mathcal{N}\,,
\label{free1}
\end{align}
where an irrelevant field-independent constant has been dropped. 
To proceed we employ the \emph{curvature approximation}
\begin{align}
\ln \mathcal{J}[A] - \ln \mathcal{J}[\ab] =  
\ln \det\big(-\hat{\db}_\mu \hat{D}^\mu(\ab + Q)\big) - \ln \det \big(-\hat{\db}_\mu \hat{\db}^\mu \big) 
\approx - \frac{1}{2}\,\int Q_\mu^c \,\chi^{cd}_{\mu\nu}\,Q_\mu^d \,,
\label{curvxx}
\end{align}
where both sides of eq.~(\ref{curvxx}) vanish at $Q=0$ and the curvature kernel $\chi$ 
is determined to give the optimal approximation of the lhs by a Gaussian (up to two-loop order).
The explicit calculation for $\chi$ leads to an integral equation that relates it to the 
ghost form factor $\eta(k)$, which in turn is coupled to the gluon propagator through its 
own DSE, cf.~eq.~(\ref{green}).  The detailed treatment of the ghost sector in our 
approach has been described in Refs.~\cite{Reinhardt:2004mm, Quandt:2013wna} and
the resulting integral equations can also be found in appendix \ref{app:chi}. 
It should be noted at this point that the expression on the lhs of eq.~(\ref{curvxx})
arises from the corresponding Landau gauge expression by the replacement 
$\mathbb{1}\,\partial_\mu \to \hat{\db}_\mu$. This will allow us to relate the curvature 
in eq.~(\ref{curvxx}) to the one in Landau gauge  further below. 

The normalization constant $\mathcal{N}$ can now be computed explicitly and we end up with 
the final form of our trial measure,
\begin{align}
\langle \cdots \rangle = \left[\det \frac{\ww}{2\pi}\right]^{-\frac{1}{2}} \int \mathscr{D} Q\,
\cdots\,\exp\left[- \frac{1}{2}\int Q\,\ww\,Q\right]\,,
\label{ansatz2}
\end{align}
where $\ww \equiv \omega + (1-2\alpha)\chi$.
This is now suitable for variation. The free action becomes 
\begin{align}
F[\ab] =  \langle S_{\rm YM}[\ab + Q] \rangle + \frac{\xi}{2}\,\left \langle \int \big(\hat{\db}_\mu\, Q_\mu\big)^2 
\right\rangle + \frac{1}{2}\,\int \ww^{-1}\,\chi - \ln \det\big(-\hat{\db}_\mu \hat{\db}^\mu\big) + 
\frac{1}{2}\ln \det \frac{\ww}{2\pi} \,.
\label{free2}
\end{align}
This is exactly the free action as found in Landau gauge \cite{Quandt:2013wna}, except for some 
shifts by the background field in strategic places. We will discuss the effect of these shifts in 
the next section.

\section{Relating background gauge and Landau gauge}
\label{relating}
Up to this point, the background field has been arbitrary.
Since we are only interested in the effective potential for the Polyakov loop, we can take the 
background field to have only a temporal component, which can be chosen \emph{constant} and 
Abelian, $\ab_\mu = \delta_{\mu 0}\, \ab$ with $\ab = \ab^{c'} T^{c'}$. We also observe that
the entire free action (\ref{free2}) including the curvature approximation (\ref{curvxx})
depends on the background field only through its (adjoint) covariant derivative
$\hat{\db} = \partial + \hat{\ab}$. This was 
already observed for the FP operator above, and it also holds for the gauge fixing term 
and the YM action because
\begin{align}
 S_{YM}[\ab + Q] \sim \tr\big(\hat{F}(\ab + Q)^2\big) = \tr \,\big[\hat{D}_\mu(\ab + Q)\,,\,
 \hat{D}_\nu(\ab + Q)\big]^2 = \tr\big[ \hat{\db}_\mu + \hat{Q}_\mu\,,\,\hat{\db}_\nu + \hat{Q}_\nu
 \big]^2\,.
\end{align}
It is therefore convenient to go to an adjoint color base in which the matrix $\hat{d}$ or 
$\hat{\ab}^{ab} = \ab^{c'} \,[T_{c'}]^{ab} = \ab^{c'} f^{ac'b}$ is diagonal. For $G=SU(2)$, 
this is simply the spherical basis introduced in Ref.~\cite{Reinhardt:2013iia}: The $(3 \times 3)$
matrix $[\hat{T}_3]^{ab} = -\epsilon^{ab3}$ has eigenvalues $\lambda = -i \sigma$ with 
$\sigma \in \{-1,0,1\}$. The corresponding eigenvectors are 
\[
\vek{e}_{\sigma=1} = - \frac{1}{\sqrt{2}} \begin{pmatrix}1 \\ i \\ 0 \end{pmatrix}\,,\qquad\qquad
\vek{e}_{\sigma=0} = \begin{pmatrix}0 \\ 0 \\ 1 \end{pmatrix}\,,\qquad\qquad
\vek{e}_{\sigma=-1} = \frac{1}{\sqrt{2}} \begin{pmatrix}1 \\ -i \\ 0 \end{pmatrix}\,.
\]
We use greek letters $\sigma, \tau\,\ldots \in \{-1,0, 1\}$ to denote color components in the new 
spherical basis, and latin letters $a,b \in {1,2,3}$ for the usual Cartesian components of adjoint 
$SU(2)$ color. 
Any matrix $M^{ab}$ in the adjoint representation can then be transformed to the new 
basis using\footnote{For simplicity, we use the same symbol for the matrices in the two bases, 
which are distinguished by having either latin or greek indices.}
\[
M^{ab} = \sum_{\sigma,\tau} \vek{e}_\sigma^a\,M^{\sigma \tau}\,(\vek{e}_\tau^b)^\ast\,.
\]
In particular, this unitary transformation diagonalises $\hat{T}^3$. The advantage of the new 
color basis is that the covariant background derivative is diagonal,
\begin{align}
\hat{\db}_\mu^{\sigma \tau} = \delta^{\sigma \tau}\,\db_\mu^\sigma\,,\qquad\qquad
\db_\mu^\sigma = \partial_\mu + \delta_{\mu 0}\,(-i \sigma \ab)\,.
\end{align}
In momentum space, i.e.~when acting on $e^{i p x}$, this operator becomes
\begin{align}
\db_\mu^\sigma(p) = i (p_\mu - \sigma \ab\,\delta_{\mu 0}) \equiv 
i  p^\sigma_\mu
\label{shift}
\end{align}
and we find the d'Alembertian
\begin{align}
[- \hat{\db}^2]^{ab} &= \vek{e}_\sigma^a (\vek{e}_\sigma^b)^\ast\,(-\Delta^\sigma)\,,
&
-\Delta^\sigma(p) &\equiv - \db_\mu^\sigma(p)\db_\mu^\sigma(p) = \vek{p}^2 + (p_0 - \sigma \ab)^2\,.
\label{alembert}
\end{align}
A similar root decomposition exists for $G=SU(3)$, where the single index 
$\sigma \in \{-1, 0, 1\}$ must be replaced by a two-component vector (the so-called 
\emph{root vector}) $\Vek{\sigma} = (\sigma_3,\sigma_8)$, because $SU(3)$ has rank 2
and hence two Cartan generators. This is worked out in appendix \ref{app:su3}. One finds
that the $SU(3)$ algebra can be decomposed into three $SU(2)$ sub-algebras, and the 
$SU(3)$ potential can thus be written as a suitable "skew" superposition of $SU(2)$ 
potentials. For simplicity, we will therefore continue to use $SU(2)$ notation until we 
present explicit $SU(3)$ results. 

Returning to our variational ansatz eq.~(\ref{ansatz2}), we note that the kernel $\ww$ cannot 
be taken color diagonal in the original (Cartesian) basis due to the background field, but it 
\emph{can} be taken diagonal in the spherical basis because this diagonalises $\hat{\db}_\mu$, 
which is the only way in which the background field enters. Furthermore, the constant background 
field does not break translational invariance and we can Fourier transform the kernel as usual, 
so that
\begin{align}
\ww^{ab}_{\mu\nu}(p) = \vek{e}^a_\sigma\,(\vek{e}_\tau^b)^\ast\,\ww^{\sigma \tau}_{\mu\nu}(p)
= \vek{e}^a_\sigma\,(\vek{e}_\tau^b)^\ast\,\delta^{\sigma \tau}\,\ww^\sigma_{\mu\nu}(p).
\end{align}
We can relate the kernel $\ww_{\mu\nu}^\sigma(p)$ to the corresponding kernel in Landau
gauge. To see this, recall that \textbf{(i)} the kernels reduce to their Landau gauge 
counterpart when the background field vanishes ($\ab = 0$), and \textbf{(ii)} the background 
field $\ab$ enters the Yang-Mills action and the Faddeev-Popov determinant only through its 
(adjoint) covariant derivative $ \hat{\db}_\mu$. Furthermore, there are no derivatives in the 
free action which are \emph{not} part of a covariant derivative $\hat{\db}_\mu$. This observation
make it intuitively clear that we only need to replace $\partial_\mu \to \hat{\db}_\mu$ to obtain 
a solution of the gap equation for $\ab \neq 0$, and this amounts to a shift in the momentum of 
the Fourier transformed kernel. We will confirm this expectation in the next section.

More precisely, the replacement 
$\partial_\mu \delta^{ab} \to \hat{\db}_\mu^{ab}$ is a matrix equation and we need to go to the 
spherical color basis in which $\hat{\db}_\mu$ is diagonal. For every  root vector $\sigma$, 
the replacement $\partial_\mu \to \hat{\db}_\mu$ amounts to the shift eq.~(\ref{shift}) of 
the momentum, and we have $\ww^\sigma_{\mu\nu}(p) = \ww_{\mu\nu}(p_\sigma)$. 
Note that this shift includes the Lorentz structure as well: $\ww_{\mu\nu}$ 
is (ordinary) 4-transversal, while $\ww^\sigma_{\mu\nu}$ is background transversal, 
i.e.~proportional to the projector
\begin{align}
t^{ab}_{\mu\nu} = \vek{e}_\sigma^a (\vek{e}_\sigma^b)^\ast\,t^\sigma_{\mu\nu}\,,
\qquad\quad
t^\sigma_{\mu\nu}(p) &\equiv \delta_{\mu\nu} - \frac{\db_\mu^\sigma(p)\,
\db_\nu^\sigma(p)}{-\Delta^\sigma(p)}\,.
\label{proja}
\end{align}
The same argument holds for the curvature, since the lhs of the defining equation
(\ref{curvxx}) contains the background field only through $\hat{\db}_\mu$. 
More details on a similar argument in Coulomb gauge can be found in Ref.~\cite{Reinhardt:2013iia}. 

It should finally be noted that the relation between Landau and background gauge derived here
only holds up to two-loop order in the free action. In higher orders, it is fairly easy to see 
that the free action must contain terms in which ordinary derivatives enter in other 
combinations than just $\hat{\db}_\mu$. Because the effective action is gauge invariant 
(in background gauge), such derivatives must act on, or appear in, gauge invariant operators 
such as $\mathsf{L}$. This was already observed in Ref.~\cite{Braun:2007bx} where it was 
also argued that such corrections are small or negligable.

\section{The effective action for the Polyakov loop}
\label{EA}
Let us go back to the expression eq.~(\ref{free2}) for the free action. To keep the 
formulars readable, we employ an obvious shorthand notation where a roman digit stands 
for the combination of spactime, Lorentz and adjoint color index, 
$1 \equiv (x, \mu,a),\,\,2 \equiv (y,\nu,b)$ etc.
The expectation values of the Yang-Mills and gauge fixing action in the Gaussian trial 
measure eq.~(\ref{ansatz2}) can be worked out using Wick's theorem. After some algebra, 
we obtain
\begin{align}
F[\ab] = &\frac{1}{2}\big[\gamma(1,2) + \chi(1,2)\big]\,\ww^{-1}(1,2) + \frac{1}{2}\,
\gamma(1,2,3,4)\,\big[\ww^{-1}(1,2)\,\ww^{-1}(3,4) + \text{(2 perm)}\big] +
\nonumber \\[2mm]
&{}+\frac{1}{2}\,\ln \det \frac{\ww}{2\pi}  + \frac{1}{2}\,\ln\det(-\hat{\db}^2)
- \ln \mathcal{J}[\ab]\,.
\label{free3}
\end{align}
Here, the kernels $\ww$ and $\chi$ are matrices in color, spacetime and Lorentz space;
as for the latter, they multiply the transversal Lorentz projector (\ref{proja}). The kernels
$\gamma$ are the bare proper vertices from the Yang-Mills and gauge fixing action,
\begin{align}
\gamma(1,2) &\equiv \gamma(x,\mu,a\,|\,y,\nu,b) = \Big[- \delta_{\mu\nu}\,\hat{\db}^2_{ab}(x) + 
(1-\xi^{-1})\,\hat{\db}_\mu^{ac}(x)\,\hat{\db}_\nu^{cb}\Big]\,\delta(x,y)
\label{gammas}\\[2mm]
\gamma(1,2,3,4) &\equiv \gamma(x,\mu,a\,|\,y,\nu,b\,|\,u,\alpha,c\,|\,v,\beta,d)
= \frac{g^2}{2}\,f^{abe}\,f^{ecd}\,\delta_{\alpha\mu}\,\delta_{\beta\nu}\,
\delta(y-x)\,\delta(u-x)\,\delta(v-x)\,. \nonumber
\end{align}
Furthermore, the second to last term in eq.~(\ref{free3}) 
is the contribution from the longitudinal degree of freedom (which receives no radiative 
correction), and the last term comes from the two perturbative ghost degrees of freedom,
$\mathcal{J}[\ab] = \mathrm{det}\big(-\hat{\db}_\mu \hat{\db}_\mu\big)$, 
because the curvature gives only the \emph{ratio} of the full and free background 
FP determinant, cf.~eq.~(\ref{curvxx}).
The last two terms in eq.~(\ref{free3}) decouple from the dynamics, but they are 
required to get the counting of the degrees of freedom right.

The optimal gluon propagator can be found by minimising eq.~(\ref{free3}) with respect 
to the variation kernel,
\begin{align}
\ww(1,2) = \gamma(1,2) + \chi(1,2) + \frac{\delta \chi(3,4)}{\delta\ww^{-1}(1,2)}\,
\ww^{-1}(3,4) + \Big[ \gamma(1,2,3,4)\,\ww^{-1}(3,4) + \text{5 perm}\Big]\,.
\label{free4}
\end{align}
We define the tadpole contraction of the 4-gluon vertex by 
\begin{align}
M_0^2(1,2) = \Big[\gamma(1,2,3,4) + \text{2 perm} \Big]\,\ww^{-1}(3,4)
= g^2\,f^{ace}\,f^{ebd}\,\delta_{\mu\nu}\,\big(\ww^{-1}\big)^{cd}_{\alpha\alpha}(x,x)\,
\delta(x,y)\,.
\label{tadpole}
\end{align}
The permutations in the free energy eq.~(\ref{free4}) can also be worked out using 
the Jacobi identity for the structure constants, and they are found to give just an 
overall factor of 2 as compared to the tadpole contraction. The result can therefore 
be written in the form
\begin{align}
F[\ab] &= \frac{1}{2}\,\Big[ \gamma(1,2) + M_0^2(1,2) + \chi(1,2)\Big]\,\ww^{-1}(1,2) + 
\frac{1}{2}\ln \det\ww + \frac{1}{2}\,\ln \det (-\hat{\db}^2)- \ln \mathcal{J}[\ab]\,,
\nonumber \\[2mm]
\ww(1,2) &= \gamma(1,2) + 2 M_0^2(1,2) + \chi(1,2) + \ww^{-1}(3,4)\,\frac{\delta \chi(3,4)}
{\delta \ww^{-1}(1,2)}\,.
\label{gapsimple}
\end{align}
This is formally identical to the expression in Landau gauge. The difference is that the 
kernel $\gamma(1,2)$ and the FP determinant (and hence the curvature $\chi$) now depend on the 
background gauge field $\ab$. More precisely, they are obtained from their Landau gauge 
counterpart by replacing each partial derivative by a background-covariant derivative,
$\mathbb{1}\,\partial_\mu \to \hat{\db}_\mu$. For a constant Abelian background field 
$\ab$, this amounts to the shift eq.~(\ref{shift}) in momentum spaces. As a consequence,
we now conclude that the solution $\ww$ of the gap equation in background gauge is also 
obtained from its Landau gauge counterpart by the same momentum shift. This is very 
beneficial as we do not have to re-solve the gap equation when varying the background field.

One remark concerning the last term in the gap equation (\ref{gapsimple}) is in order.
In Refs.~\cite{Quandt:2013wna} and \cite{Quandt:2015aaa}, we have simply discarded this 
term as a higher loop effect. A more systematic scheme would introduce a formal 
loop counting parameter $\lambda$ as a prefactor in the exponent of the trial measure 
(\ref{ansatz2}), which amounts to counting (fully dressed) internal gluon lines. 
In this scheme, our results (including the curvature approximation) are correct to 
order $\mathcal{O}(\lambda)$ in the gap equation, and $\mathcal{O}(\lambda^2)$ in the 
free energy. However, it is then intuitively clear that the last term in the gap 
equation (\ref{gapsimple}) is of the same order as the previous one and cannot be 
discarded. More precisely, a direct calculation (cf.~appendix \ref{app:chi}) 
reveals that 
\begin{align}
\chi(1,2) = \chi_0 + \omega^{-1}(3,4)\, \frac{\delta \chi(3,4)} {\delta \ww^{-1}(1,2)}
+ \text{2-loop}\,,
\label{fischer}
\end{align}
where $\chi_0$ is a field- and momentum-independent quadratic divergence that is 
eventually removed by renormalization.\footnote{It would entirely be absent if we used 
e.g.~dimensional regularization.} Since the tadpole term $M_0^2$ is a similar
quadratic divergence, it is prudent to combine the two in the form
\begin{align}
\bar{\chi}(1,2) \equiv \chi(1,2) + M_0^2(1,2)\,.
\label{combi}
\end{align}
This results in the free action and gap equation, respectively, 
\begin{align}
F[\ab] &= \frac{1}{2}\,\Big[ \gamma(1,2) + \xx(1,2)\Big]\,\ww^{-1}(1,2) + 
\frac{1}{2}\ln \det\ww + \frac{1}{2}\,\ln \det(-\hat{\db}^2)- \ln \mathcal{J}[\ab]\,,
\nonumber \\[2mm]
\ww(1,2) &= \gamma(1,2) + 2 \xx(1,2) - \chi_0\,.
\label{gapcool}
\end{align}
Inserting the solution of the gap equation into $F[\ab]$ yields the (unrenormalized) 
self-consistent effective action of the background field $\ab$,
\begin{align}
\Gamma[\ab] = -\frac{1}{2} \big[\chi(1,2)-\chi_0\big]\,\ww^{-1}(1,2) +
\frac{1}{2}\ln \det\ww + \frac{1}{2}\,\ln \det(-\hat{\db}^2)- \ln \mathcal{J}[\ab]\,.
\label{R1}
\end{align}
The subtraction $(\chi-\chi_0)$ in the first term already removes the quadratic divergence 
in the curvature, and the sole effect of the renormalization is to further cancel the 
subleading logarithmic divergence in the curvature. 

To see how this comes about, we need to introduce counter terms. The gap equation 
only requires a gluon mass and field counter term, while a third (ghost wave function) 
counter term is necessary for the ghost form factor which enters the integral equation for the 
curvature.\footnote{The ghost sector is worked out in detail in Refs.~\cite{Quandt:2013wna} 
and \cite{Quandt:2015aaa}.} The renormalized gap equation thus reads
\begin{align}
\ww(1,2) &= \gamma(1,2) + 2 \xx(1,2) - \chi_0 + \delta Z_A\,\gamma(1,2) + \delta M^2\,.
\label{gapren}
\end{align}
The mass counter term $\delta M^2$ serves to cancel the quadratic divergence $\chi_0$ in 
the curvature ($\delta M^2 + \chi_0 = 0)$, while the subleading logarithmic divergence is 
compensated by the wave function renormalization $\delta Z_A$. (The finite pieces in the 
counter terms are fixed by precise renormalization conditions on the gluon propagator 
as described in Ref.~\cite{Quandt:2015aaa}.) With these arrangements, we find
\begin{align}
\ww(1,2) &= \gamma(1,2) + 2 \xx(1,2) + 2 \,\delta M^2 + \delta Z_A\,\gamma(1,2) 
\nonumber \\[2mm]
&\equiv Z \,\gamma(1,2) + \chi_R(1,2)\,,
\label{truegap}
\end{align}
were $Z$ is a finite part of $\delta Z_A$ determined by the renormalization condition
$\ww(\mu) = Z\,\mu^2$ at a large scale $\mu \gg 1$, cf.~next section. Note that the 
renormalized curvature in the gap equation
\begin{align}
\chi_R(1,2) \equiv \xx(1,2) + \delta M^2 + \frac{1}{2}\,\Big[\delta Z_A + 1 - Z\Big]\,\gamma(1,2)\,, 
\label{chiR}
\end{align}
is unambiguously determined by the counter terms. This would not have been possible if we 
had discarded the last term in eq.~(\ref{gapsimple}), as we could not combine the 
curvature and tadpole contribution in this case. Since only that combination has a 
unique (mass) counter term, we would then be faced with the problem of how to "distribute"
the finite parts of the mass counter term onto $M_0^2$ and $\chi$ to define these quantities 
individually. 

The renormalized free action using the definition eq.~(\ref{chiR}) is
\begin{align}
F[\ab] = \frac{1}{2}\,\Big[ Z \gamma(1,2) + \chi_R(1,2) \Big]\,\ww^{-1}(1,2) + 
\frac{1}{2}\ln \det\ww + \frac{1}{2}\,\ln \det(-\hat{\db}^2)- \ln \mathcal{J}[\ab]\,.
\end{align}
If we finally employ the renormalized gap equation (\ref{truegap}), we find the final expression for 
the self-consistent effective action of background field,
\begin{align}
 \Gamma[\ab] = - \frac{1}{2}\,\chi_R(1,2) \,\ww^{-1}(1,2) + \frac{1}{2}\ln \det\ww + 
 \frac{1}{2}\,\ln \det(-\hat{\db}^2)- \ln \mathcal{J}[\ab]\,.
 \label{gold}
\end{align}
This is the renormalized version of eq.~(\ref{R1}). The apparent logarithmic 
divergence in $\Gamma[\ab]$ drops out when taking the \emph{difference} to the 
Landau case of vanishing background field, $\Gamma[\ab]-\Gamma[0]$. This is the 
(finite) quantity we are heading for.

\bigskip
To put equation (\ref{gold}) in a managable form, a few more steps are required. 
First, we write out the roman digit notation in position space and employ 
translational invariance (for a constant background field) to Fourier transform to 
momentum space. Let us first write down the result for Landau gauge $\ab = 0$:
\begin{align}
 \Gamma[0] = \frac{1}{2}\,(N^2-1)\,V_4\int \frac{d^4k}{(2\pi)^4}\,
 \Bigg[- 3 \,\frac{\chi_R(k)}{\ww(k)} + 3 \,\ln \frac{\ww(k)}{2\pi} + 
 \ln \frac{\omega_{\|}(k)}{2 \pi} - 2 \ln k^2\Bigg]\,.
 \label{brezel}
\end{align}
Here, the spacetime volume $V_4$ has factorized because of translational invariance, and 
we are really computing the effective \emph{potential} $\Gamma / V_4$.
Furthermore, the variational kernel $\ww(k)$ (and likewise the curvature) is 
color diagonal and transversal in Landau gauge, $\ww_{\mu\nu}^{ab}(k) = 
\delta^{ab}\,t_{\mu\nu}(k)\,\ww(k)$ with a \emph{scalar} variation kernel $\ww(k)$. 
The color trace yields the prefactor $(N^2-1)$, while the Lorentz trace gives the 
factors of $3$ for the transversal degrees of freedom, one for the longitudinal 
degree of freedom, and $(-2)$ for the ghost degrees of freedom. Since the curvature
describes the deviation from the free-field FP determinant only, we must also 
include the contribution from the free ghost degrees of freedom $\mathcal{J}[0]$, 
which is the last term in eq.~(\ref{brezel}). Furthermore, the longitudinal gluon 
receives no radiative corrections beyond one loop and we have $\omega_{\|}(k) = k^2$.

In the next step, we must put the system at finite temperatures since we want to study 
the deconfinement phase transition. This amounts to imposing periodic boundary 
conditions along the compactified euclidean time direction of length $\beta$, which 
is the inverse temperature. As a consequence, the momentum integrals are 
\[
\int_\beta \dd k\,f(k) = \beta^{-1}\sum_{n \in \mathbb{Z}} \int \frac{d^3 k}{(2\pi)^3}\,f(\nu_n, 
\mathbf{k})\,,
\]
i.e.~the integral over $k_0$ is always understood as a discrete sum over the Masubara frequencies
$k_0 = \nu_n = 2 \pi n / \beta$.  Moreover, the heat bath singles out a rest frame 
and the overall $O(4)$ invariance of the theory is broken. As a consequence, the gluon propagator 
(and also the curvature) comes in two \emph{distinct} Lorentz structures, which are both 
4-dimensionally transversal, but also 3-dimensionally longitudinal or transversal, respectively. 
The corresponding kernels  $\ww_\perp$ and $\ww_\|$ (and $\chi_\perp$, $\chi_\|$) were computed 
in Ref.~\cite{Quandt:2015aaa}: as the temperature increases, there is a moderate suppression 
of the gluon propagators, and a slight enhancement of the ghost form factor. The temperature 
sensitivity is larger in the components longitudinal to the heat bath, but this affects only 
the infrared region.  Generally, the ghost and gluon propagators are only moderatley changed 
by temperatures well up to twice the critical temeprature $2 T^\ast$, and we can even discard 
their implicit temperature dependence and use the zero-temperature kernels throughout. 
On the one hand, this reduces the 
numerical effort dramatically, since the finite temperature propagators are easily 
three orders of magnitude harder to calculate.\footnote{The reason is that the $T=0$ model 
can use $O(4)$ invariance to reduce the momentum integral to a double integral over the 
momentum norm and one angle. The same can be done for the spatial momentum integral at 
finite temperature, which is thus of the same complexity. On top of this, however, 
the finite temperature solution must sum up to $40-50$ Matsubara frequencies which 
represent a coupled channel problem that must be solved by iteration. This easily 
adds three orders of magnitude to the numerical effort.} 
It is also justified \emph{a posteriori} by our numerical results, since the dominant 
contribution to the (Poisson-resummed) Matsubara series comes from frequencies in the 
mid-momentum regime around $1\,\text{GeV}$, where the Green's functions are only very 
slightly affected by finite temperature. As long as we are only interested in the 
demonstration of the basic phenomenon without excessive accuracy goals, the use of the 
$T=0$ propagators is thus justified. 

\bigskip
We have seen above that the (zero-temperature) propagators in Landau gauge can be 
re-used in the presence of a constant Abelian background field. All we have to do is 
to replace the color trace factor $(N^2-1)$ in eq.~(\ref{gold}) by a sum over all 
root vectors, and for each root $\sigma$ shift the momentum  argument in the kernels 
by 
\begin{align}
k_0 = \frac{2 \pi}{\beta}\,n \to \frac{2\pi}{\beta}\,n + \sigma \ab_0^3 = 
\frac{2 \pi}{\beta}(n + \sigma \,x)\,,
\qquad\qquad x \equiv \frac{\beta \ab_0^3}{2 \pi} \in [0,1]\,.
\end{align}
(This is for $G=SU(2)$; we will discuss $G=SU(3)$ below.) The root $\sigma=0$ 
gives the unshifted contribution ($\ab = 0$) and hence drops out from the 
physically relevant difference $\Gamma[\ab]-\Gamma[0]$.
Since the kernels do not depend on the sign of $k_0$, positive and negative roots
give equal contributions and we can restrict ourselves to $\sigma=1$ and 
include a factor of two. After factorizing the 4-volume and performing the 
integral over the angles, we end up with
\begin{align}
\beta^4\,V_{\rm eff}(x) &= 12 \pi \int_0^\infty dq\,q^2\,
\sum_{n \in \mathbb{Z}}\,\Bigg[\left(\ln \frac{\bar{\omega}(k_n(x))}{\bar{\omega}(k_n(0))}  
- \ln \frac{k_n(x)^2}{k_n^2}\right) -\Big(\frac{\chi_R(k_n(x)) }{\bar{\omega}(k_n(x))}  
- \frac{\chi_R(k_n(0))}{\bar{\omega}(k_n(0))}\Big) + 
\nonumber \\[2mm]
& \hspace*{5cm} + \frac{2}{3}\ln \frac{k_n(x)^2}{k_n^2}\Bigg]
\label{basis}
\end{align}
Here, $q \equiv |\vek{k}| \beta / (2\pi)$ is the rescaled dimensionless momentum and 
\begin{align}
k_n(x) \equiv \frac{2\pi}{\beta}\sqrt{(n+x)^2 + q^2}
\end{align}
is the norm of the shifted 4-momentum. In the first term  in eq.~(\ref{basis}), we 
have subtracted the perturbative contribution ($\ww(k) = k^2$) for the three transversal 
gluons, and added it back in the last term, combined with similar contributions from 
the longitudinal gluon and the perturbative ghost. These contributions in the last term
of eq.~(\ref{basis}) are known as the \emph{Weiss potential} \cite{Weiss:1980rj}, and it 
can be calculated explicitly (see appendix \ref{app:poisson}), 
\begin{align}
\beta^4\,W(x) \equiv 8 \pi \int_0^\infty dq\,q^2\,\sum_{n \in \mathbb{Z}}\,
\ln \frac{(n+x)^2 + q^2}{n^2 + q^2} =  \frac{4}{3}\,\pi^2 x^2 (1-x)^2\,.
\end{align}
Even with the Weiss potential subtracted, eq.~(\ref{basis}) is not well suited for numerical 
evaluation. Instead, it is prudent to first do some arithmetic manipulations such as 
Poisson resummation, introduction of spherical 4-coordinates etc. This  is presented 
in detail in appendix \ref{app:poisson}. The result is the equivalent formula
\begin{align}
 \beta^4\,V_{\rm eff}(x) &= \beta^4 W(x) + \frac{6}{\pi^2} \sum_{m=1}^\infty 
 \frac{1 - \cos(2 \pi m x)}{m^4}\,h(\beta m) \nonumber \\[2mm]
 h(\lambda) &= - \frac{1}{4} \int_0^\infty d\xi\,\xi^2 J_1(\xi)\,\Bigg[ 
 \ln\left[\frac{\bar{\omega}(\xi/\lambda)}{(\xi/\lambda)^2}\right]
 - \frac{\chi_R(\xi/\lambda)}{\bar{\omega}(\xi/\lambda)} \Bigg]\,,
 \label{baz}
\end{align}
where $J_1$ is a regular Bessel function and $\xi = \lambda k = m \beta k$ is dimensionless.
We will use this formula as the basis for our numerical investigation, with the curvature 
from eq.~(\ref{chiR}) and the inverse gluon propagator $\ww(k)$ from the renormalized 
$T=0$ gap equation (\ref{truegap}), see also Ref.~\cite{Quandt:2015aaa} for details.

\section{Numerical setup and results}
\label{numerics}
We use eq.~(\ref{baz}) directly to compute the effective action of the Polyakov loop 
for $G=SU(2)$. The integral over the rescaled momentum $\xi$ is tricky because of the 
oscillating nature of the Bessel function. Since the argument of $J_1(\xi)$ is temperature
independent (which was the purpose of the rescaling $q \to \xi$), we can precompute the 
roots of $J_1$ and break up the integral at these roots into many small contributions
from each half-oscillation of $J_1$. The remainder of the integrand is positive, and so
the contributions come with alternating sign and a series accelerator can be used to 
accurately estimate the remainder. We have checked spot values against \textsc{Mathematica} 
to ensure that this procedure is reliable. The remaining series over the frequencies 
$m$ converges quite well due to the factor $1/m^4$. It can also be estimated by an 
accelerator, or simply summed up to convergence.

\begin{figure}[t]
\centering
\includegraphics[width=7cm,keepaspectratio=true]{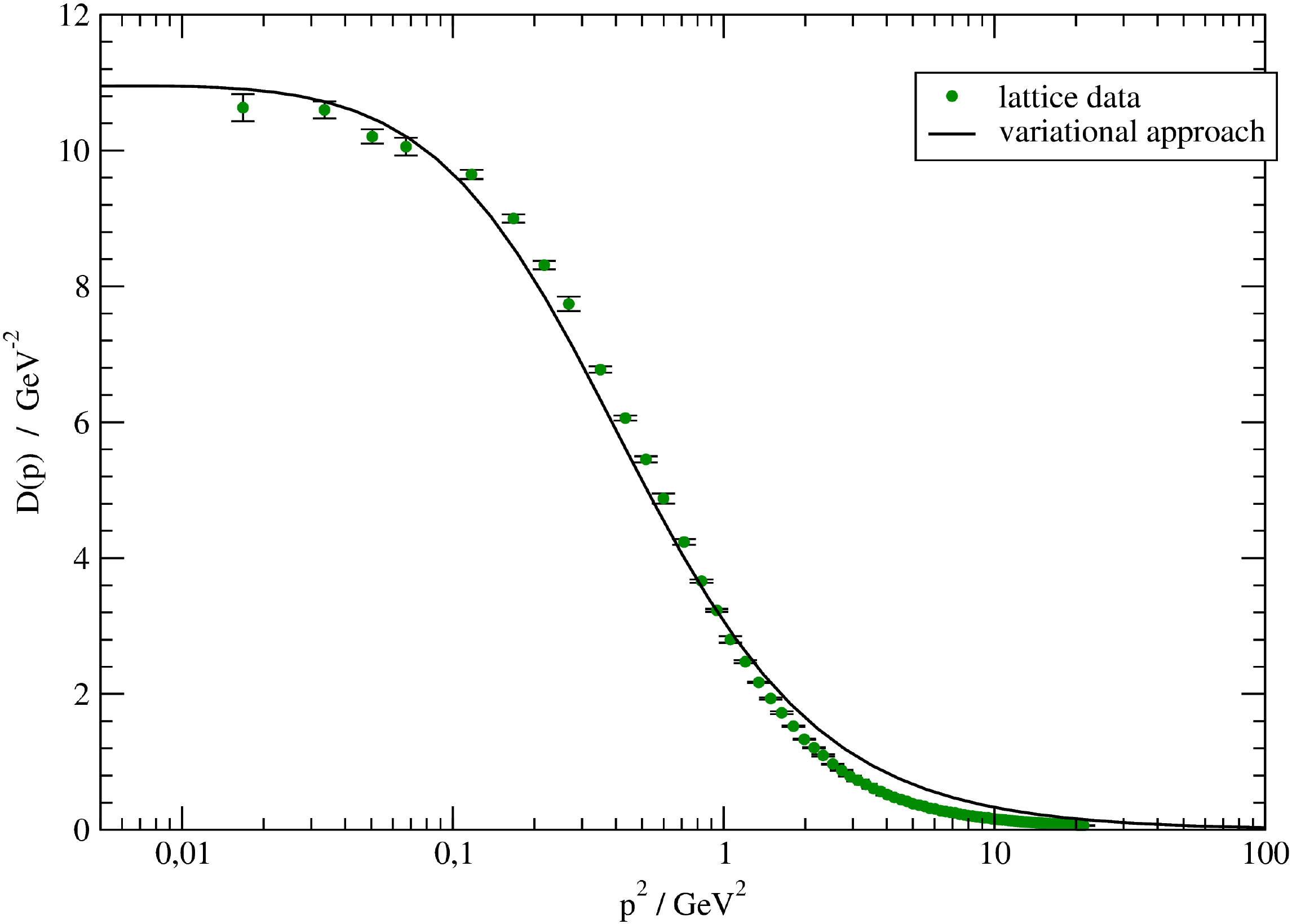}
\hspace{1cm}
\includegraphics[width=7cm,keepaspectratio=true]{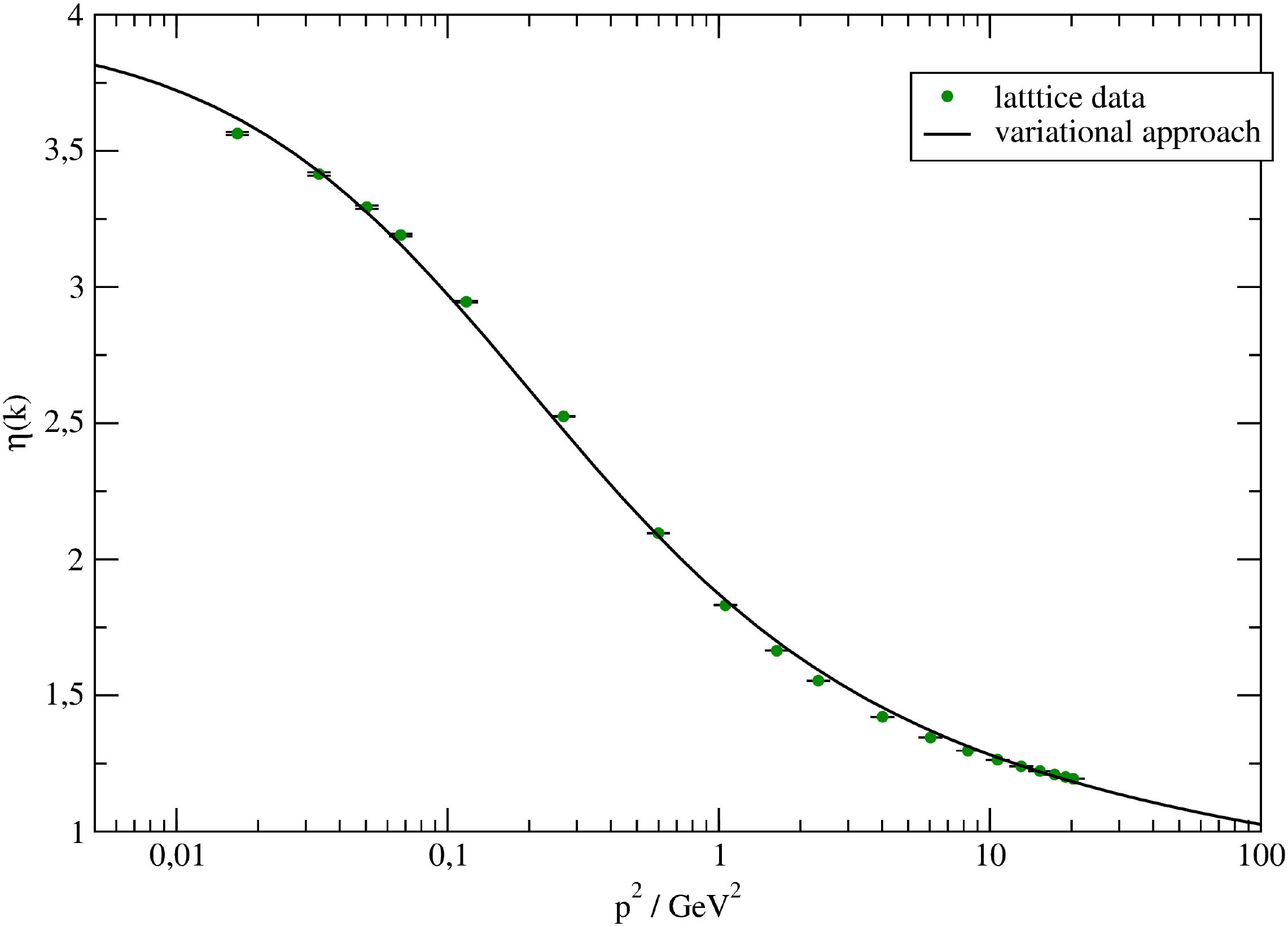}
 \caption{ \label{fig:0} The zero-temperature gluon propagator (\emph{left}) and 
 the ghost form factor (\emph{right}) in Landau gauge computed from the 
 improved gap equation eq.~(\ref{truegap}), and compared against high-precision 
 lattice data taken from Ref.~\cite{Bogolubsky:2009dc}.}
\end{figure}

For the inverse gluon propagator $\ww(k)$, we take the numerical $T=0$ solution of the 
gap equation (\ref{truegap}), with the renormalization conditions as discussed in 
Ref.~\cite{Quandt:2015aaa}. Since we no longer discard the derivative term in  
eq.~(\ref{gapsimple}), the gap equation itself differs slightly from the 
one considered in Ref.~\cite{Quandt:2015aaa}. In a first step, we must therefore 
recompute the $T=0$ solution in Landau gauge and determine the renormalization 
constants which best fit the lattice data. We employ the same high-precision 
lattice data \cite{Bogolubsky:2009dc} used for  Refs.~\cite{Quandt:2013wna, Quandt:2015aaa}
and leave the renormalization conditions unchanged. The solutions for the 
improved gap equation eq.~(\ref{truegap}) with the renormalized curvature taken 
from eq.~(\ref{chiR}) (and the integral equations for $\chi(k)$ and the ghost 
form factor $\eta(k)$ as in Ref.~\cite{Quandt:2015aaa}, see also appendix \ref{app:chi}) 
is presented in Fig.~\ref{fig:0}. 
As can be seen, the improved gap equation describes the lattice 
propagators with the same (or even slightly better) accuracy than in the previous 
studies. The optimal value for the renormalizaltion scale $\mu$ is slightly larger 
than the $\mu=5\,\mathrm{GeV}$ used previously; this also affects the absolute 
value of the gluon mass parameter. The renormalization points are $\mu_c = 0$ 
for the ghost, $\mu_0 = 113\,\mathrm{MeV}$ for the gluon mass parameter, and 
$\mu = 5.64\,\mathrm{GeV}$ which also determines the overall scale. The best 
values for the renormalization parameters are 
\begin{align}
\eta(0)^{-1} = 0.2533\,,\qquad\quad Z = 0.3127\,,\qquad\quad M_A = 541\,\mathrm{MeV}\,.
\label{params}
\end{align}
At this scale, the coupling constant comes out as $N g^2 = 4.64$ or 
$\alpha \equiv g^2 / (4 \pi) = 0.19$, 
but this is not an adjustable parameter, as it is determined uniquely by eq.~(\ref{params}).
The solution in Fig.~\ref{fig:0} is initially optimised for the color group
$SU(2)$ and the renormalization constants could change slightly when optimizing 
against $SU(3)$ lattice data. Within our truncation scheme, the number $N$ of colors 
appears only in the effective coupling $Ng^2$, which can be re-adjusted by 
rescaling the propagators and correcting the gluon mass parameter $M_A$.
Preliminary fits indicate, however, that the optimal mass parameter  $M_A$ 
for $G=SU(3)$ is not much different from the best value for $G=SU(2)$, and this 
has an even minor effect on the Polyakov loop potential studied here. Within 
the accuracy attempted in this study, it is thus sufficient to use the same 
solution for the $T=0$ gluon and ghost propagator in the remaining numerical 
study.

\subsection{$G=SU(2)$}

For the computation of the Polyakov loop potential, we use eq.~(\ref{baz}) with the 
(inverse) gluon propagator determined at $T=0$ within our formulation as described above.  
It should be emphasized that all (three) renormalization constants are fixed at $T=0$ 
and there are \emph{zero} adjustable parameters for the entire rest of the calculation,
i.e. the complete family of Polyakov loop potentials, the \emph{vev} of the Polyakov loop
itself and, in particular, the phase transition temperatures $T^\ast$. The lattice data 
is also not used in the calculation itself, except for the initial determination of the 
appropriate renormalization constants in our approach at $T=0$. 

\begin{figure}[t]
\centering
\includegraphics[width=7cm,keepaspectratio=true]{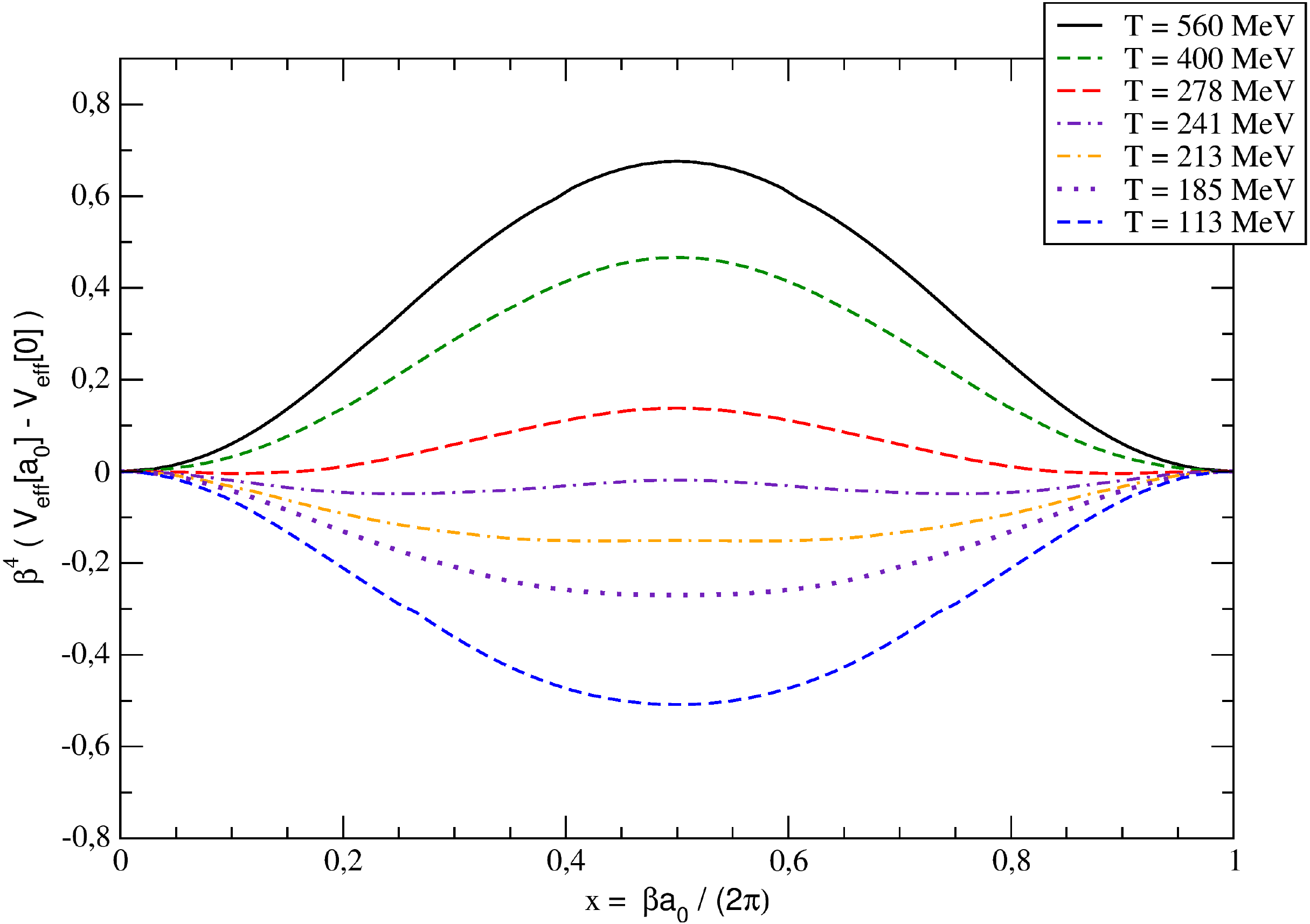}
\hspace{1cm}
\includegraphics[width=7cm,keepaspectratio=true]{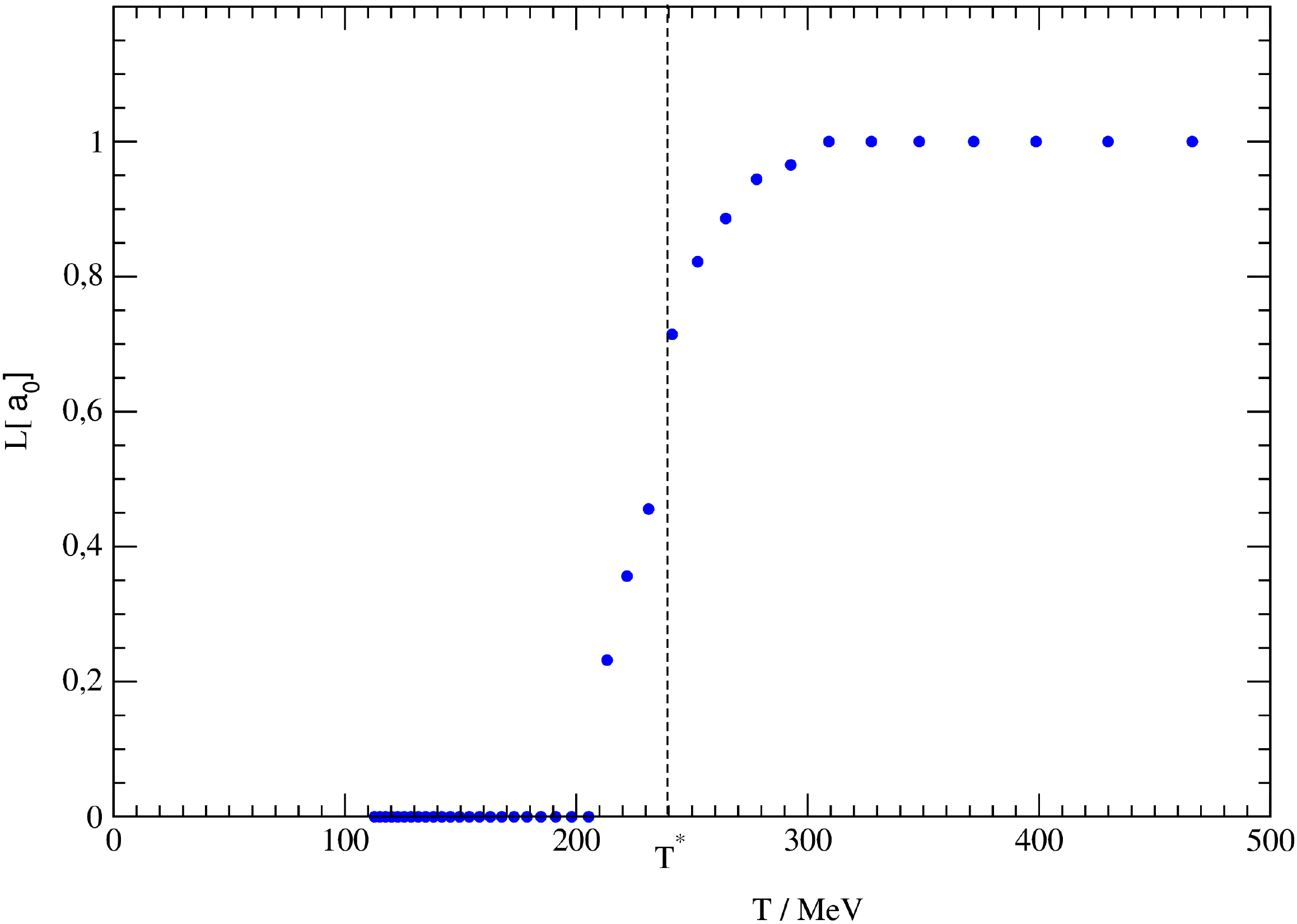}
 \caption{[color online]  \label{fig:1} \emph{Left}: The effective potential $V_{\rm eff}(x)$ 
 for the rescaled $SU(2)$ Polyakov loop $x = \beta \ab_0^3 / (2\pi) \in [0,1]$ at various 
 temperatures, decreasing from top to bottom.
 \emph{Right}: The traced $SU(2)$ Polyakov loop $\mathsf{L}(\langle A_0 \rangle)$ 
 as a function of temperature.}
\end{figure}

In the left panel of Fig.~\ref{fig:1}, we plot the effective potential $V_{\rm eff}(x)$ 
for the Polakov loop $x = \beta \ab_0^3 / (2\pi)$ on the fundamental domain $x\in[0,1]$ for 
various temperatures. We find a clear phase transition at the point where the minimum 
$V_{\rm eff}$ moves away from the center-symmetric value $\bar{x}=\frac{1}{2}$.
From the location $\bar{x}$ of the minimum of $V_{\rm eff}$, we find the preferred background 
field $\ab_0^3 = 2 \pi \bar{x} / \beta$ and hence the Polyakov loop 
$\langle \mathsf{L} \rangle \approx \mathsf{L}(\langle A_0\rangle) = \cos(\pi\,\bar{x})$.
This is plotted in the right panel of Fig.~\ref{fig:1}, from which it is apparent that the 
phase transition is \emph{second order}, in agreement with the known lattice results. The 
transition into the fully center-broken deconfined phase with
$\langle \mathsf{L} \rangle = 1$ is rather wide (which is also supported by lattice 
calculations), and we find a phase transition temperature\footnote{Because of the rather wide 
transition, the exact location of $T^\ast$ is somewhat arbitrary. We choose the 
point with the maximal slope of $\langle \mathsf{L} \rangle$ as a function of $T$, 
i.e.~the peak of the suszeptibility.}
\begin{align}
T^\ast \approx 239\,\mathrm{MeV} \qquad\qquad (G=SU(2))\,. 
\end{align}
This should be compared to the lattice result of $T^\ast = 312\,\mathrm{MeV}$
determined
in Ref.~\cite{Lucini:2003zr} by a careful finite size scaling analysis.
In view of the simplicity of our Gaussian ansatz, this is a reasonable numerical agreement. 
The quality of the numerical accuracy could be further improved by 
taking the actual finite temperature propagators instead of the $T=0$ solution, or 
by enlarging our ansatz space (using non-Gaussian measures, dressed vertices etc.). This would,
however, complicate the analysis and, in particular, make the numerical evaluation 
much more costly. The main purpose of this paper is, instead, to show that a relatively
simple picture of a constitutent gluon coupled to an infrared enhanced ghost 
can describe the physics of the deconfinement transition in all qualitative aspects, 
and even in fair quantitative agreement to the lattice.

\subsection{$G=SU(3)$}
This group has rank 2 and the effective potential $V_{\rm eff}(x,y)$ is thus a function of 
two parameters which are the rescaled Cartan components of the background field
\begin{align}
x = \frac{\beta \,\ab_0^3}{2\pi} \in [0,1]\,,\qquad \qquad 
y = \frac{\beta \,\ab_0^8}{2 \pi} \in \big[0, \frac{2}{\sqrt{3}}\big]\,.
\label{dah}
\end{align}
The potential must be computed as a sum over the $SU(2)$ potentials corresponding to 
the three non-trivial positive root vectors. The appropriate shift in the momentum $p_\sigma$ 
is determined in detail in appendix \ref{app:su3}. 

\begin{figure}[t]
\centering
\includegraphics[width=7cm,keepaspectratio=true]{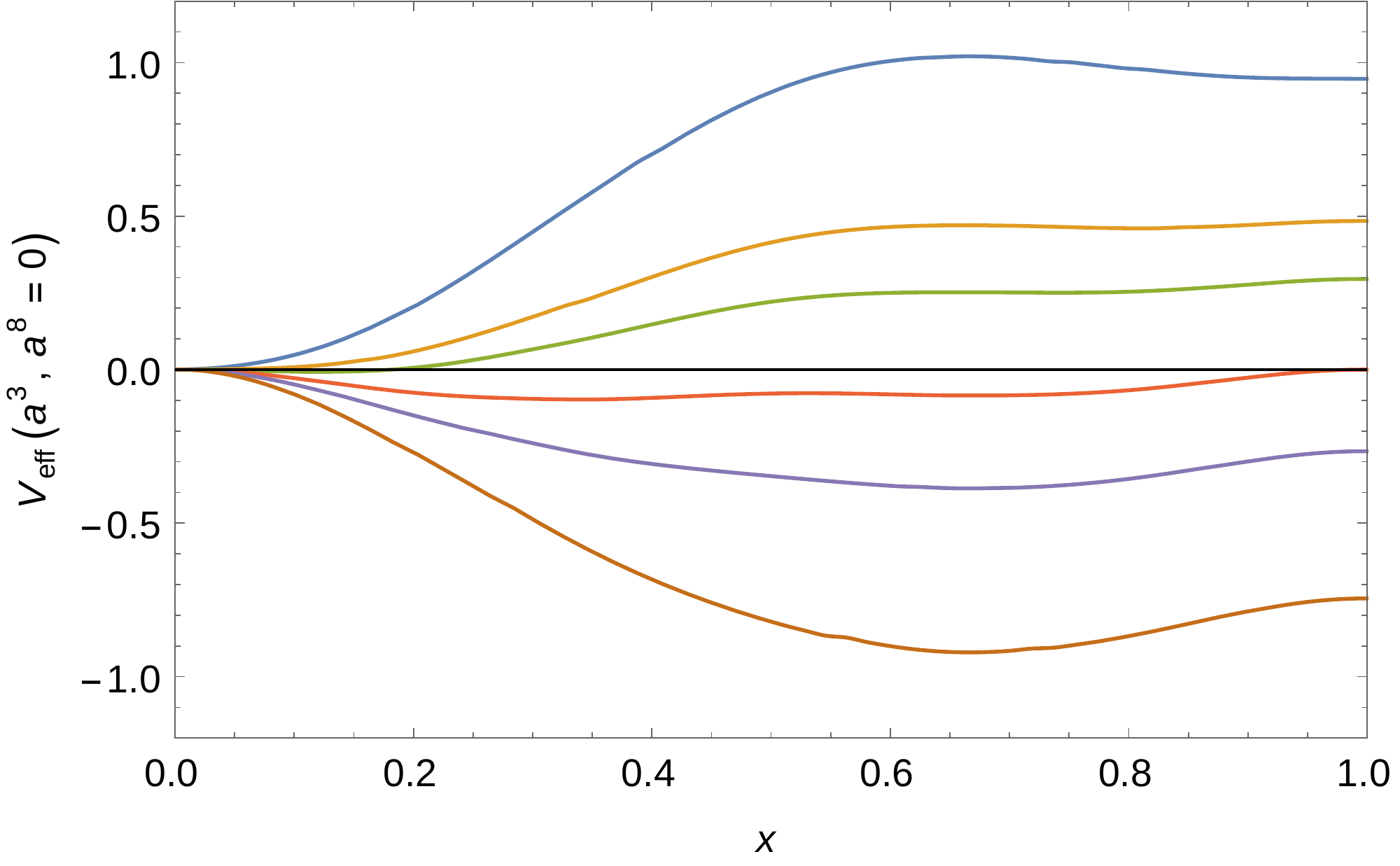}
\hspace{1cm}
\includegraphics[width=7cm,keepaspectratio=true]{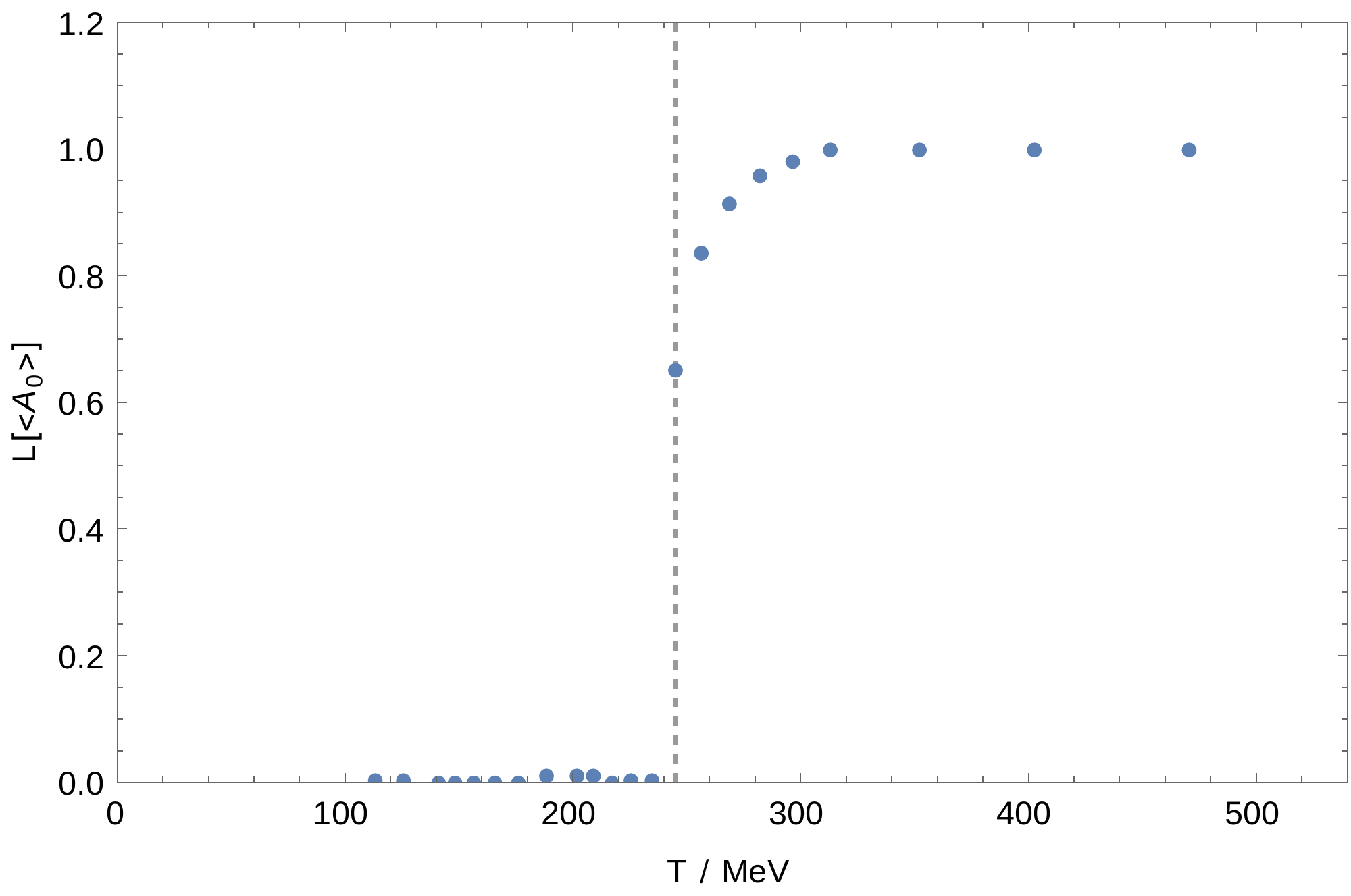}
 \caption{[color online] \label{fig:3}\emph{Left}: A slice of the effective potential $V_{\rm eff}(x, y=0)$ 
 for the color group $G=SU(3)$ at various temperatures, decreasing from 
 $403\,\mathrm{MeV}$ (top) to $166\,\mathrm{MeV}$ (bottom).
 \emph{Right}: The traced $SU(3)$ Polyakov loop $|\mathsf{L}(\langle A_0 \rangle)|$ 
 as a function of temperature.}
\end{figure}

In the left panel of Fig.~\ref{fig:3} we have plotted the slice $y=\ab_0^8 = 0$ 
of the effective potential as a function of $x \in [0,1]$ for several temperatures. 
The potential is no longer symmetric around $x=\frac{1}{2}$, because the center 
symmetric points (where $\mathsf{L}[\ab] = 0$) are at different locations 
in the Weyl alcove
\begin{align}
\text{center symmetric } (x,y)\,:\qquad \big(\frac{2}{3}\,,\,0\big)\,,\qquad \quad
\big(\frac{2}{3}\,,\,\frac{2}{\sqrt{3}}\big)\,,\qquad \quad 
\big(\frac{1}{3}\,,\,\frac{1}{\sqrt{3}}\big)\,.
\label{sky}
\end{align}
By contrast, the center breaking minima of the Weiss potential are positions
where the traced Polyakov loop is one of the center elements,
$\mathsf{L}[\ab] \in \{ 1, e^{2 \pi i / 3},\, e^{-2 \pi i / 3}\}$ or 
\begin{align}\,
\text{center broken } (x,y)\,:\qquad \big(0\,,\,0\big)\,,\qquad \quad
\big(0\,,\,\frac{2}{\sqrt{3}}\big)\,,\qquad \quad 
\big(1\,,\,\frac{1}{\sqrt{3}}\big)\,.
\label{dumont}
\end{align}
By center symmetry, the are always three degenerate minima of the effective 
potential, which all give the same absolute value for the Polyakov loop $\mathsf{L}$, 
cf.~Figs.~\ref{fig:4} and \ref{fig:5} below.
In the right panel of Fig.~\ref{fig:3}, we have plotted the value of 
$|\mathsf{L}|$ at the minima as 
a function of temperature. We now observe a phase transition that is clearly 
\emph{first order}, in accordance with lattice findings. Our best estimate for 
the phase transition temperature is 
\begin{align}
T^\ast = 245\,\mathrm{MeV}\qquad\qquad(G=SU(3))\,.
\end{align}
This should be compared to the lattice estimate \cite{Lucini:2003zr} of 
$T^\ast \approx 284 \,\mathrm{MeV}$. Again, we observe that the qualitative features 
of the deconfinement phase transition (such as its order) are correctly predicted, and
the numerical estimate of the transition temperature is in reasonable agreement with 
the lattice data. The accuracy of the agreement is actually better than could be expected, 
since we have not bothered to re-optimise the renormalization constants of our approach 
for the case of $SU(3)$, i.e.~the entire calculation is still based on $SU(2)$ propagators.
This is also one obvious way for numerical improvement, in addition to what was already 
suggested for $G=SU(2)$ above.  In view of this, the good numerical agreeement with the 
lattice may be somewhat accidental.

\begin{figure}[t]
\centering
\includegraphics[width=6.5cm,keepaspectratio=true]{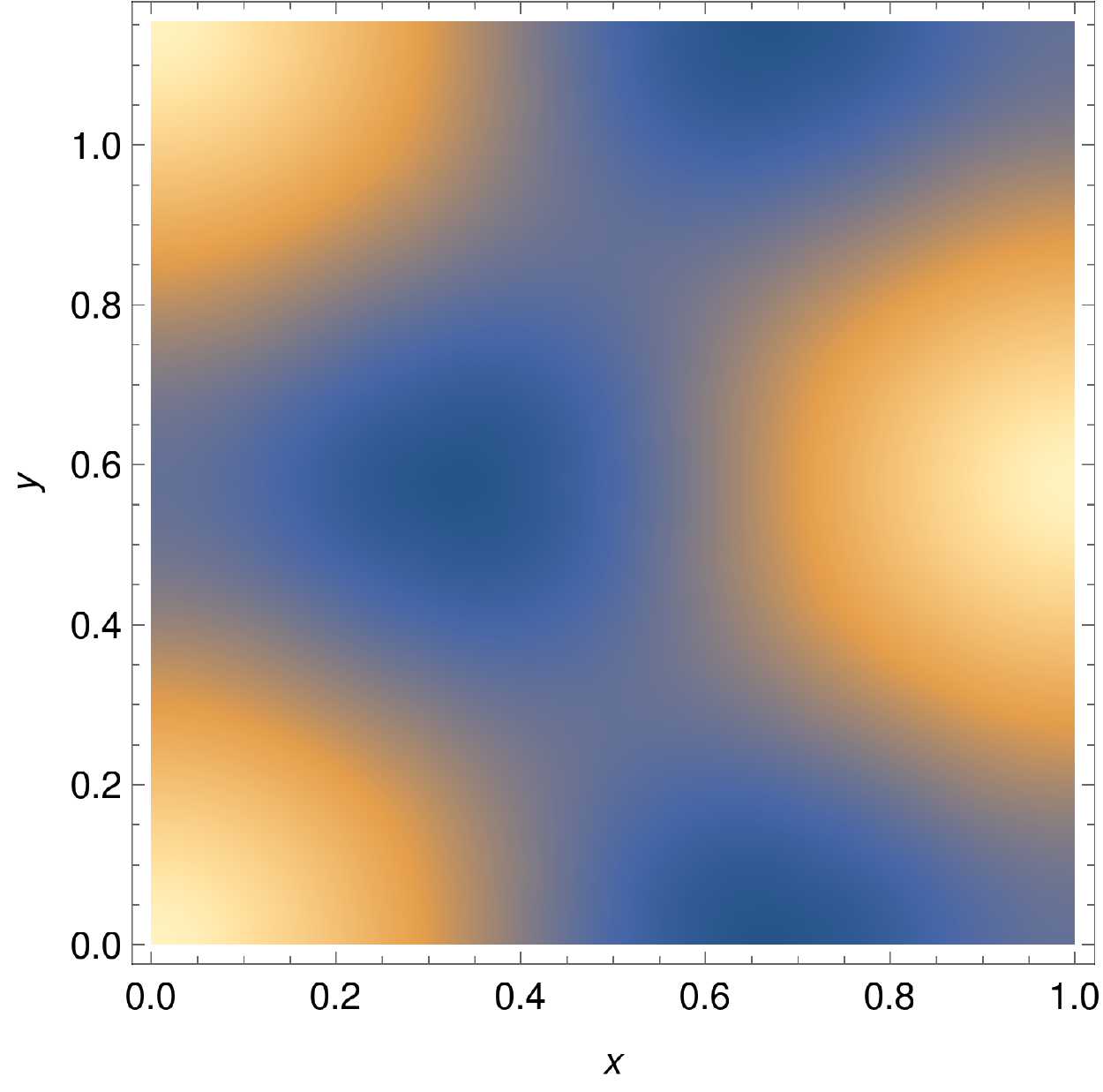}
\hspace{1cm}
\includegraphics[width=7.5cm,keepaspectratio=true]{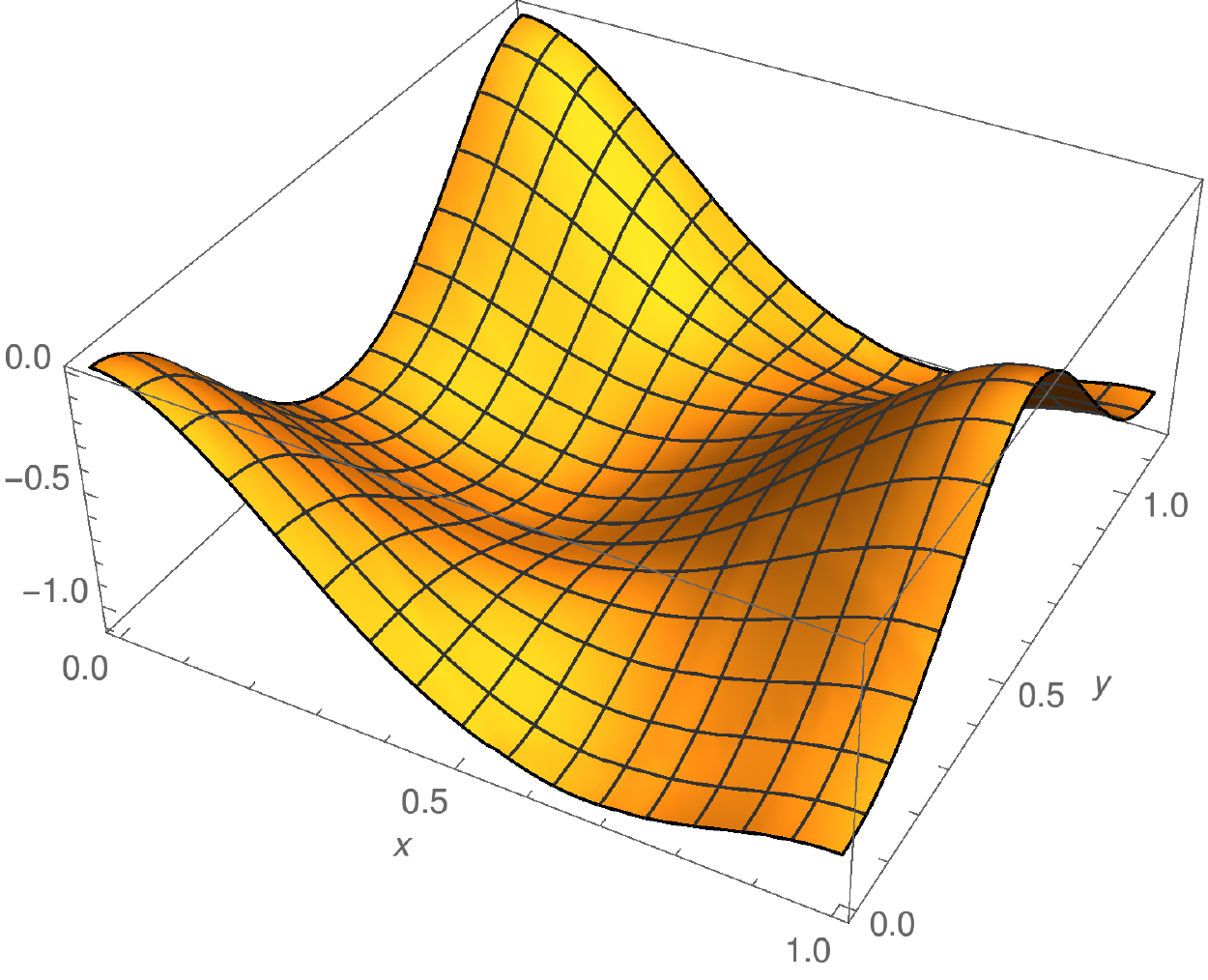}
 \caption{[color online] \label{fig:4} The effective Potential $V_{\rm eff}(x,y)$ of 
 the $SU(3)$ Polyakov loop over the fundamental domain, computed at a temperature 
 $T = 141\,\mathrm{MeV}$ in the confined phase. In the density plot on the left, 
 darker color shades indicate a lower value of the effective potential. Both plots
 clearly demonstrate  that the minima occur at the center symmetric points eq.~(\ref{sky}) 
 where the Polyakov loop vanishes, i.e.~center symmetry is preserved.}
\end{figure}

To better visualize the Polyakov loop potential $V_{\rm eff}(x,y)$, we have also plotted
it as a function of both $SU(3)$ Cartan parameters. Figure \ref{fig:4} shows the 
result for a single temperature $T= 141\,\mathrm{MeV}$ below the phase transition, 
i.e.~in the confined phase. Both the surface and density plot clearly indicate that the minima 
occur at the center symmetric points (\ref{sky}) with $\mathsf{L} = 0$. By contrast, 
Fig.~\ref{fig:5} shows the same plot for a temperature $T = 400\,\mathrm{MeV}$ above the 
phase transition, i.e.~in the confined phase. Now the minima are clearly at the 
symmetry breaking points eq.~(\ref{dumont}) with $|\mathsf{L}| = 1$.

\begin{figure}[t]
\centering
\includegraphics[width=6.5cm,keepaspectratio=true]{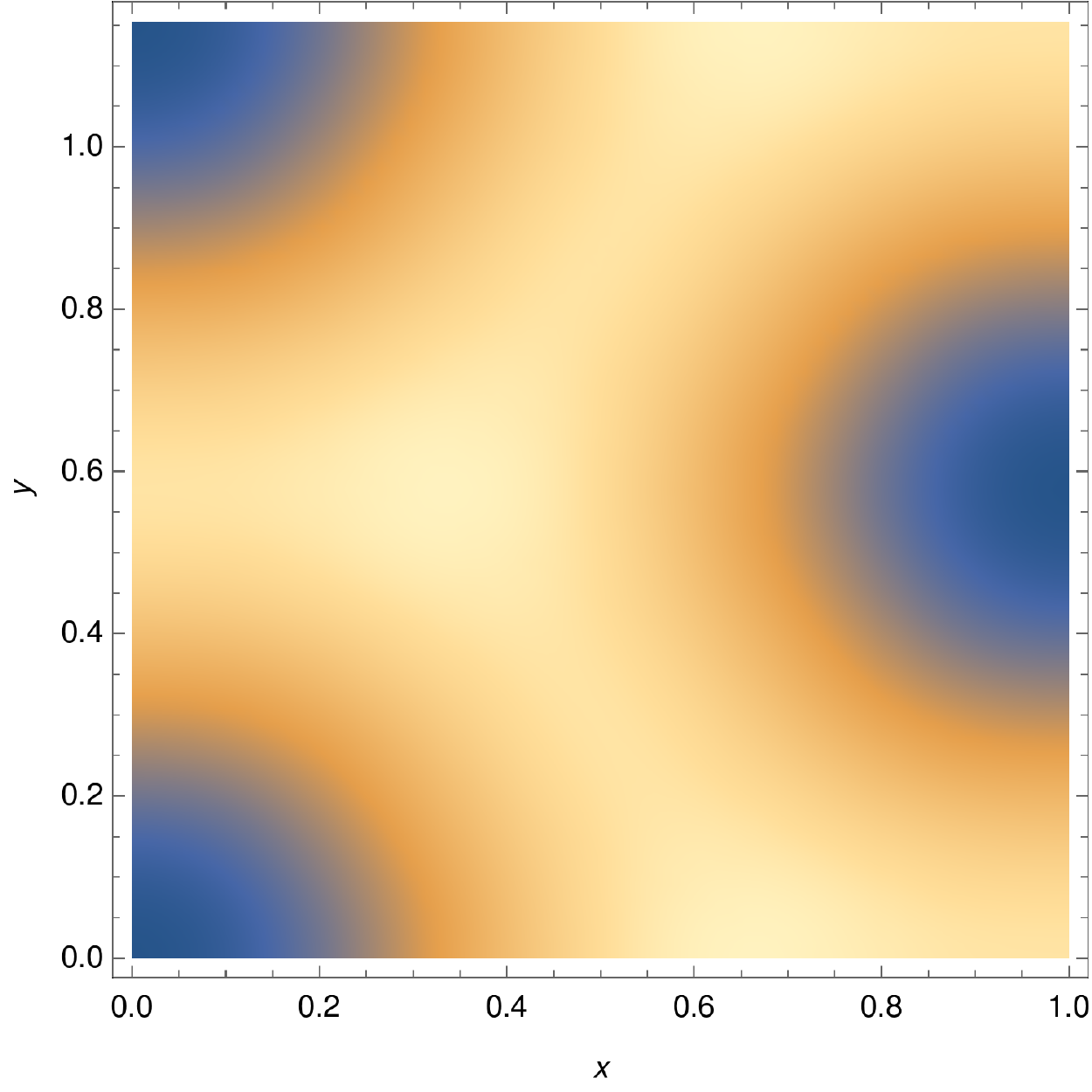}
\hspace{1cm}
\includegraphics[width=7.5cm,keepaspectratio=true]{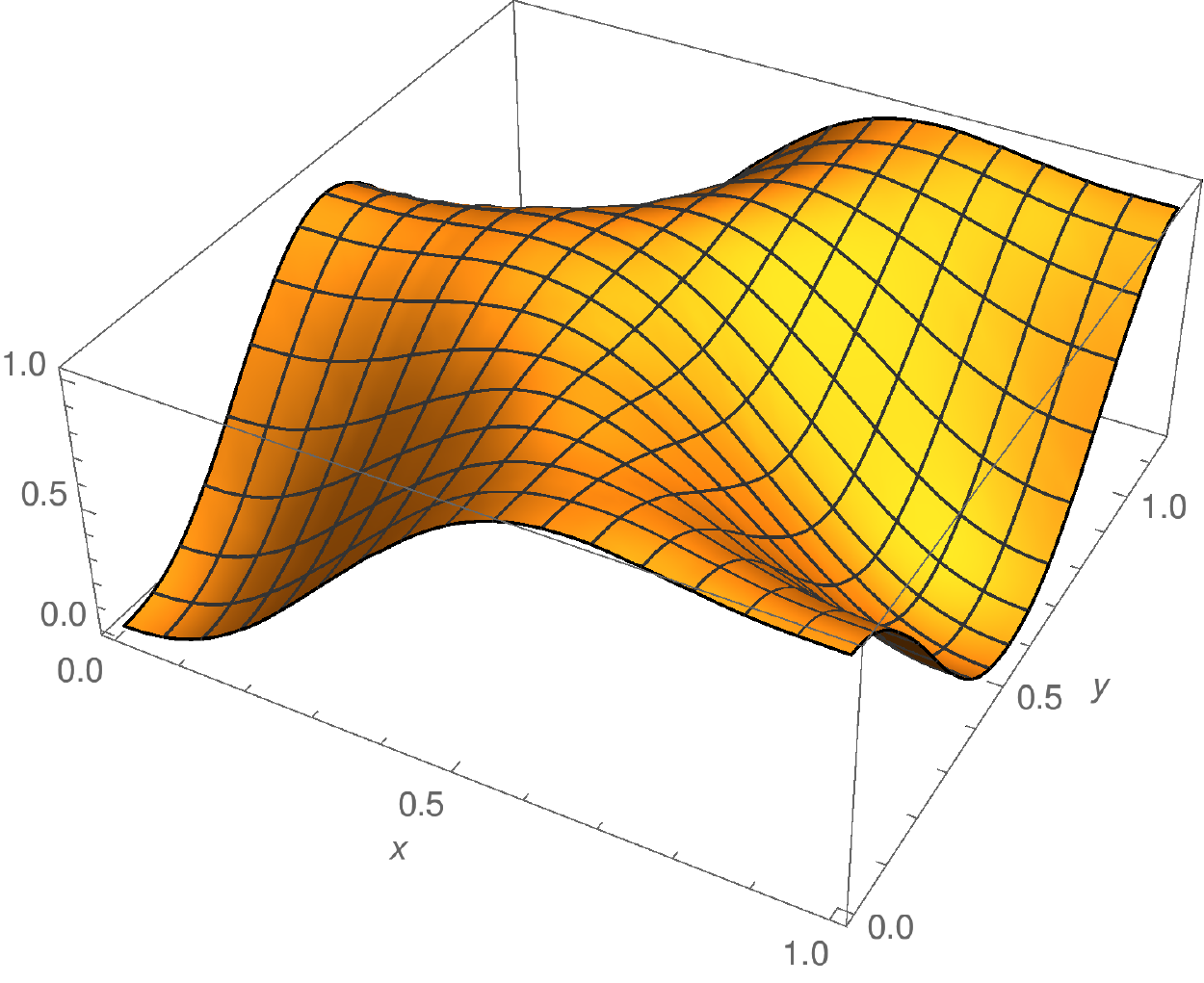}
 \caption{[color online]\label{fig:5} The effective Potential $V_{\rm eff}(x,y)$ 
 of the $SU(3)$ Polyakov loop over the fundamental domain, computed at a temperature 
 $T = 400\,\mathrm{MeV}$ in the deconfined phase. This should be compared to  
 Fig.~\ref{fig:4}: The minima now occur at the points eq.~(\ref{dumont}) where the 
 Polyakov loop is a center element, and center symmetry is consequently broken.}
\end{figure}

\section{Conclusions}
\label{conclusions}
In this paper, we have extended the covariant variational approach to Yang-Mills 
theory to background gauge and studied the effective action for the Polyakov loop.
Our findings demonstrate that the formalism is not only well suited to describe 
low-order Green's functions, but it can also accurately capture the physics of 
the deconfinement phase transtition. All known qualitative ascpects of this 
transition (in particular its order for different color groups) are correctly 
reproduced, and there is even reasonable numerical agreement with the transition
temperatures measured on the lattice. Given the simplicity of our ansatz without
adjustable parameters except for the $T=0$ renormalization constants, this agreement 
is quite remarkable. 
Based on these encouraging findings, it is planned to extend the present formulation 
to include fermions and study the QCD phase diagram also in the region of 
non-vanishing chemical potentials. Preliminary studies in this  direction 
are currently underway.

\begin{acknowledgments}
This work was supported by \emph{Deutsche Forschungsgemeinschaft} 
(DFG) under contract DFG-RE856/9-2.
\end{acknowledgments}


\appendix

\section{Curvature}
\label{app:chi}
In this appendix, we want to show eq.~(\ref{fischer}) in the main text. We start from 
the Dyson-Schwinger equation for the ghost form factor in Landau gauge,
\begin{align}
\eta(k)^{-1} = 1 - Ng^2\,\int \dd q\,\big[1-(\hat{k}\cdot\hat{q})^2\big]\,\frac{\eta(k-q)}
{(k-q)^2\,\ww(q)}\,.
\end{align}
If we emplyo the loop counting scheme explained before eq.~(\ref{fischer}), this can 
be solved in the form
\begin{align}
\eta(k) = 1 + Ng^2\,\int \dd q\, \big[1-(\hat{k}\cdot \hat{q})^2\big]\,\frac{\eta^{[1]}(k-q)}
{(k-q)^2\,\ww(q)}\,\lambda + \mathcal{O}(\lambda^2)
\label{blue}
\end{align}
and we find
\begin{align}
\frac{\delta \eta(k)}{\delta \ww^{-1}(p)} = \lambda\,Ng^2 \,  \big[1-(\hat{k}\cdot\hat{p})^2\big]
\,\frac{\eta^{[1]}(k-p)}{(k-p)^2}\,\lambda + \mathcal{O}(\lambda^2)\,.
\label{red}
\end{align}
Next we consider the integral equation for the curvature,
\begin{align}
\chi(k) = \frac{Ng^2}{d-1}\,\int \dd q\,\big[1- (\hat{k}\cdot\hat{q})^2\big]\,\frac{\eta(k-q)\,\eta(q)}
{(k-q)^2}\,.
\label{green}
\end{align}
Taking the derivative w.r.t.~$\ww^{-1}(p)$ under  the integral and using eq.~(\ref{red}) yields, after 
some algebra,
\begin{align}
\int \dd p\,\ww^{-1}(p)\,\frac{\delta \chi(k)}{\delta \ww^{-1}(p)} = \lambda\,
\frac{Ng^2}{d-1}\,\int \dd q\,\frac{1 - (\hat{k}\cdot\hat{q})^2}{(k-q)^2}\,\Big[ \eta^{[1]}(k-q) 
+ \eta^{[1]}(q)\Big]\,.
\end{align}
This is precisely the $\mathcal{O}(\lambda)$ contribution if the expansion eq.~(\ref{blue}) is 
used in the curvature eq.~(\ref{green}). The term of order $\mathcal{O}(\lambda^0)$ is
\begin{align}
\chi_0 = \frac{Ng^2}{3}\,\int \dd q\,\big[1-(\hat{k}\cdot \hat{q})^2\big]\,\frac{1}{(k-q)^2}
= - \frac{2\,Ng^2}{105 \pi^3}\cdot \Lambda^2 = \text{const}\,\Lambda^2\,,
\end{align}
where the numerical prefactor was computed in $d=4$ using an $O(4)$-invariant momentum cutoff. 
(Other schemes will give different numerical prefactors.) Thus, we have shown
\begin{align}
 \chi(k) = \chi_0 + \lambda\cdot \int \dd p\,\ww(p)^{-1}\,\frac{\delta \chi(k)}{\delta \ww^{-1}(p)} + 
 \mathcal{O}(\lambda^2)\,.
\end{align}
This is the momentum space representation of eq.~(\ref{fischer}) in the main text. 
Although the derivation was carried out explicitly for Landau gauge, it generalizes to the 
case of a constant Abelian background field, since the necessary shift in the momentum 
arguments has no consequence at this point.

\section{Root decomposition of $SU(N)$}
\label{app:su3}
The semi-simple Lie algebra $SU(N)$ has rank $r=(N-1)$ and there are hence 
$r$ mutually commuting generators $H_k$ which span the Cartan subalgebra
of $SU(N)$. As explained in the main text, the background field 
$\ab_\mu = \ab\,\delta_{\mu 0}$ must be chosen in the Cartan subalgebra,
\[
\ab = \sum_{k=1}^r \ab^k \,H_k\,. 
\]
Since the $H_k$ are anti-hermitean and mutually commuting, they can be 
simultaneously diagonalized with purely imaginary eigenvalues$(-i \mu_k)$.
The real numbers $\mu_k$ are called the weights of $H_k$, and the collection
of one eigenvalue from each $H_k$ forms a weight vector 
$\Vek{\mu} = (\mu_1,\ldots,\mu_r)$. The number of such vectors, i.e.~the number 
of the eigenvalues of $H_k$ depends on the representation. 

In the present paper, we are mainly concerned with the background field in
the \emph{adjoint} representation, $\hat{\ab}^{ab} = - f^{abc}\,\ab^c$. The weights 
$\sigma_k$ in the adjoint representation are called the roots,
\begin{align}
\hat{H}_k\,|\sigma\rangle = - i \sigma_k \,\vert\,\sigma\rangle
\label{HH}
\end{align}
and the real numbers $\sigma_k$ from all Cartan generators are collected in 
root vectors $\Vek{\sigma} = (\sigma_1,\ldots,\sigma_r)$. Since $\hat{H}_k$ is 
$(N^2-1) \times (N^2-1)$ dimensional, there can be at most $N^2-1$ root vectors.
However, $r=(N-1)$ must vanish and the entire root system of $SU(N)$ contains
$N(N-1)$ non-vanishing  vectors. They can be given a partial ordering by the
first element, i.e.~the eigenvalues of $\hat{H}_1$. Then the non-vanishing roots 
$\Vek{\sigma}$ come in pairs and half of the roots are positive, half of them 
are negative. From eq.~(\ref{HH}), the adjoint 
background field is diagonal in the basis $|\sigma\rangle$, 
\[
\hat{\ab}\,| \sigma \rangle = - i (\mathsf{a}\cdot\Vek{\sigma})\,| \sigma \rangle
= -i \left(\sum_{k=1}^r \ab_k\,\sigma_k\right)\,\vert \sigma \rangle\,.
\]
This is the analogon to the cyclic $SU(2)$ basis used in the main text. 
The complete argument about the diagonalization for $\hat{\db}_\mu$ and the 
shift $p \to p - p_\sigma$ in the momentum to go from Landau to background 
gauge remains valid, if only $p_\sigma$ is generalized to 
\begin{align}
p^\mu \to p^\mu - p_\sigma^\mu = p^\mu - (\ab\cdot \Vek{\sigma})\,\delta^{\mu 0}\,. 
\end{align}

For $G=SU(2)$, the root and weight vectors are pure numbers since the Cartan 
subalgebra is $r=1$-dimensional. There are two weights $\pm \frac{1}{2}$ and three
roots $\{-1,0,1\}$, of which only two are non-vanishing. 

This structure easily generalizes to $G=SU(3)$, which has rank $r=2$. The two 
Cartan generators are usually taken as $H_1 = T_3 = \lambda_3 / (2i)$ and 
$H_2 = T_8 = \lambda_8 / (2i)$ in terms of Gell-Mann matrices. The root and 
weight vectors are both $r=2$-dimensional. In  fact, the weights read
\[
\Vek{\mu}\,:\qquad \Big( 0 \,,\, \frac{1}{\sqrt{3}} \Big)\,,
\qquad\quad \Big(\frac{1}{2} \,,\, \frac{1}{2 \sqrt{3}}\Big)\,,
\qquad\quad \Big( \frac{1}{2} \,,\, - \frac{1}{2\sqrt{3}}\Big)\,.
\]
More important are the $N^2-1 = 8$ root vectors, of which $N(N-1)=6$ are 
non-vanishing. As they come in pairs with opposite sign of $\sigma_1$, 
there are three non-vanishing positive roots
\begin{align}
\Vek{\sigma}\,:\qquad \Big( 1 \,,\, 0 \Big)\,,
\qquad\quad \Big(\frac{1}{2} \,,\, \frac{1}{2} \sqrt{3}\Big)\,,
\qquad\quad \Big( \frac{1}{2} \,,\, - \frac{1}{2}\sqrt{3}\Big)\,.
\label{ro}
\end{align}
From these roots, it is clear that any $SU(3)$ background field 
in the Cartan algebra, $\ab = \ab^3\,T^3 + \ab^8 T^8$ 
can conveniently be described by the rescaled components
\begin{align}
x = \frac{\beta \,\ab^3}{2\pi}\,,\qquad \qquad 
y = \frac{\beta \,\ab^8}{2 \pi}\,.
\end{align}
The fundamental domain (Weyl alcove) in these variables is given by 
\begin{align}
x \in \big[0,\,1\big]\,,\qquad\qquad 
y \in \big[0,\, \frac{2}{\sqrt{3}}\big]\,.
\label{alk}
\end{align}
Finally, the momentum shift $p_\sigma$ for the three positive roots is
\begin{align}
\Vek{\sigma}  = \big(0\,,\,1\big)\,&:\qquad \qquad 
p_\sigma = \frac{2\pi}{\beta}\,x \nonumber \\[2mm]
\Vek{\sigma}  = \big(\frac{1}{2}\,,\,\frac{1}{2}\,\sqrt{3}\big)\,&:\qquad \qquad 
p_\sigma = \frac{\pi}{\beta}\,\left(x + \sqrt{3}\,y\right) \nonumber \\[2mm]
\Vek{\sigma}  = \big(\frac{1}{2}\,,\,-\frac{1}{2}\,\sqrt{3}\big)\,&:\qquad \qquad 
p_\sigma = \frac{\pi}{\beta}\,\left(x - \sqrt{3}\,y\right)\,.
\end{align}
This will be used in the main text.

\section{Reformulation of the Matsubara sum}
\label{app:poisson}
If the kernels entering the effective potential have full $O(4)$-invariance 
(e.g.~because they are taken at $T=0$), then the effective potential for the 
Polyakov loop can be put in the following general form,
\begin{align}
 \beta^4 V_{\rm eff}(x) \equiv 12 \pi \int_0^\infty dq\,q^2\,\sum_{n \in \mathbb{Z}}
 \Bigg[f\big(2 \pi \beta^{-1} q_n(x)\big) - f\big(2 \pi \beta^{-1} q_n(0)\big)\Bigg]\,.
 \label{generic}
\end{align}
This is for $G=SU(2)$ and the argument $x = \beta \ab_0^3 / (2\pi) \in [0,1]$.
Furthermore, $q_n(x) = \sqrt{(n+x)^2 + q^2}$, and $f(k)$ is a function of 
the 4-momentum norm $k = \sqrt{k_\mu k_\mu}$. For instance, 
\begin{align*}
\text{Weiss potential} &: \quad f(k) = \frac{2}{3}\,\ln k^2 \\[2mm]
\text{Massive gluon propagator} &: \quad f(k) = \ln(k^2 + \mu^2) \\[2mm]
\text{Eq.~(\ref{basis})} &: \quad f(k) = \ln \ww(k) - \frac{\chi(k)}{\ww(k)} - \frac{1}{3}\,\ln k^2\,.
\end{align*}
To evaluate eq.~(\ref{generic}), we start by Poisson resumming the Matsubara series
\begin{align*}
\beta^4 V_{\rm eff}(x) = 12 \pi \int_0^\infty dq\,q^2\int_{-\infty}^\infty dz\,
\sum_{m \in \mathbb{Z}}e^{2 \pi i m z}\,\Bigg[f\big(2 \pi \beta^{-1}\,\sqrt{(z+x)^2 + q^2}\big) - 
f\big(2 \pi \beta^{-1}\,\sqrt{z^2 + q^2}\big)\Bigg].
\end{align*}
Next we shift the $z$-integral by introducing $s \equiv (z+x)$ and combine the terms 
in the series with index $m$ and $(-m)$. After interchanging the order of summation and 
integration,
\begin{align*}
\beta^4 V_{\rm eff}(x) = - 12 \pi \sum_{m=1}^\infty \Big[1 - \cos(2 \pi m x)\Big]
\int _0^\infty dq\,q^2 \,\int_{-\infty}^\infty ds\,e^{2 \pi i m s}\,f\big(2 \pi \beta^{-1}
\sqrt{s^2 + q^2}\big).
\end{align*}
The $q$-dependent terms in the integrand are even in $q$ so that we can extend the $q$-integral
to all of $\mathbb{R}$. For the resulting double integral 
in $(q,s)$ over $\mathbb{R}^2$, we use polar coordinates $(r,\varphi)$,
\begin{align*}
\beta^4 V_{\rm eff}(x) = - 12 \pi \sum_{m=1}^\infty \Big[1 - \cos(2 \pi m x)\Big]
\int_0^\infty dr\,r\int_0^{2\pi} d\varphi\,r^2 \,\sin^2 \varphi\cdot e^{2 \pi i m r \cos \varphi}
\cdot f(2 \pi r / \beta)\,.
\end{align*}
The $\varphi$-integral leads to 
\[
\int_0^{2 \pi} d\varphi\,\sin^2 \varphi \cdot e^{2 \pi i m r \cos \varphi} = 
\frac{J_1(2 \pi m r)}{m r }\,,
\]
where $J_1$ is a regular Bessel function. Finally, we rescale the integration variable 
in the remaining integral $r \to \xi \equiv 2 \pi m r$ and collect all pieces. The result
is 
\begin{align}
\beta^4 V_{\rm eff}(x) = - \frac{3}{2\pi^2}\sum_{m=1}^\infty \frac{1 - \cos(2 \pi m x)}{m^4}
\,\int_0^\infty d\xi\,\xi^2 \,J_1(\xi)\,f\big(\frac{\xi}{\beta m}\big)\,.
\end{align}
For later convenience, we write this formula as 
\begin{align}
 \beta^4 V_{\rm eff}(x) &= 12 \pi \int_0^\infty dq\,q^2 \sum_{n \in \mathbb{Z}}
 \Bigg[f\big(\frac{2\pi}{\beta}\,q_n(x)\big) - f\big(\frac{2\pi}{\beta}\, q_n(0)\big)\Bigg]
 \nonumber \\[2mm]
 &= \frac{6}{\pi^2}\sum_{m=1}^\infty \frac{1 - \cos(2 \pi m x)}{m^4}\,h(\beta m)
 \nonumber \\[2mm]
 h(\lambda) &= - \frac{1}{4}\int_0^\infty d\xi\,\xi^2 \,J_1(x)\,f(\xi / \lambda)\,.
 \label{joda}
\end{align}
Note that the temperature only enters through the factor $h(\lambda=\beta m)$, 
and the series now converges much more quickly due to the $1/m^4$ term. 
(For implementation details, see section \ref{numerics}.) 

To test this formula, let us take the case of a massive (transversal) gluon, 
$f(k) = \ln (k^2 + \mu^2)$. We use the proper-time representation of the logarithm,
\[
h(\lambda) = \frac{1}{4}\int\limits_{\Lambda^{-2}}^\infty \frac{ds}{s}
\int_0^\infty d\xi\,\xi^2 \,J_1(\xi)\cdot e^{-s (\mu^2 + \xi^2 / \lambda^2)} - 
\frac{1}{4}\int_0^\infty d\xi \,\xi^2 \,J_1(\chi)\,\gamma_E\,,
\]
where $\Lambda\to \infty$ is a cutoff to be lifted at the end of the calculation,
and $\gamma_E$ is Euler's constant. The second term is $\lambda$-independent and 
only leads to a temperature-independent constant in the potential, which can be dropped. 
For the first term, we can do the $\xi$-integral to obtain 
\[
h(\lambda) = \frac{\lambda^4}{16}\int\limits_{\Lambda^{-2}}^\infty \frac{ds}{s^3}\,
e^{-\lambda^2 / (4s) - s \mu^2}\,.
\]
After changing variables $s \to t = \frac{\lambda^2}{4s}$, the cutoff can be lifted,
\begin{align*}
 h(\lambda) =& \int\limits_0^{(\lambda \Lambda)^2 / 4}dt\,t\,e^{-t - (\lambda \mu)^2 / (4t)}
 \quad \stackrel{\Lambda\to \infty}{\longrightarrow} \quad 
\int_0^\infty dt\,t\,\exp\left[-t - \left(\frac{\lambda \mu}{2}\right)^2 \cdot \frac{1}{t}\right]
= \frac{1}{2}\,(\lambda \mu)^2\,K_2(|\lambda \mu|)\,,
\end{align*}
where $K_2$ is an irregular modified Bessel function. We can put this in the main formula
(\ref{joda}) and find 
\begin{align}
\beta^4 V_{\rm eff}(x) = \frac{3}{\pi^2} \sum_{m=1}^\infty \frac{1 - \cos(2 \pi m x)}{m^4}
\cdot (\beta \mu)^2\cdot K_2(m\,\beta \mu)\,.
\end{align}
The rhs depends on the temperature only in the combination $(\mu \beta)$, i.e.~increasing 
the mass is equivalent to decreasing the temperature, and vice versa. The Weiss potential
follows by including a prefactor\footnote{Instead of 3 transversal gluon modes,
we really have 3 transversal + 1 longitudinal - 2 ghost = 2 massless modes.} 
$2/3$ and sending the mass $\mu \to 0$. From 
$(\beta \mu)^2\,K_2(\beta \mu \, m) \to 2/m^2$ in this limit, we obtain eventually 
the $SU(2)$ Weiss potential \cite{Weiss:1980rj}
\begin{align}
\beta^4 W(x) = \frac{4}{\pi^2}\,\sum_{m=1}^\infty \frac{1 - \cos (2\pi m x)}{m^4}
= \frac{4}{3}\,\pi^2\,x^2 \,(1-x)^2\,.
\end{align}

\bibliographystyle{apsrev4-1}
\bibliography{polybib}
\end{document}